%% file: paper.tex
\documentclass[journal]{IEEEtran}

\usepackage{amsmath}
\usepackage{mathtools}
\usepackage{amsfonts}
\usepackage{amsthm}
\usepackage{amssymb}
\usepackage{mleftright}
\usepackage{bm}
\usepackage{bbm}
\usepackage{enumitem}
\usepackage{multirow}
\usepackage{booktabs}
\usepackage{url}
\usepackage{color}
\usepackage{algorithmic}
\usepackage[textsize=tiny]{todonotes}

\usepackage{graphicx}
\usepackage{pgf, tikz, pgfplots}
\usetikzlibrary{shapes, calc, patterns.meta}
\usetikzlibrary{shapes,arrows.meta, calc, positioning}
\pgfplotsset{compat=newest}

\newcommand\ebno{E_{\text{s}}/N_0} %

\def\R{{\mathbb R}}

\def\Mike{L}
\def\NT{N_\text{T}}
\def\NR{N_\text{R}}
\def\NTc{\tilde{N}_\text{T}}
\def\NRc{\tilde{N}_\text{R}}

\def\ebno{E_{\text{b}}/N_0}

\usepackage[ruled,vlined, linesnumbered]{algorithm2e}
\let\oldnl\nl%
\SetKwInput{KwInit}{Initialization}
\newcommand{\nonl}{\renewcommand{\nl}{\let\nl\oldnl}}%

\SetCommentSty{mycommfont}

\newcommand{\algphase}[1]{%
\tcp{\textbf{#1}}%
}

\input{kit_colors}
\colorlet{M2N32_color}{kit-cyan}
\colorlet{M2N32_color_overlay}{kit-royalblue}
\colorlet{M2N32_color_multifuncrs}{kit-blue}

\colorlet{M4N32_color}{kit-lightgreen}
\colorlet{M4N32_color_overlay}{kit-lightgreen!30!black}
\colorlet{M4N32_color_multifuncrs}{kit-green}

\colorlet{M4N16_color}{kit-orange}
\colorlet{M4N16_color_overlay}{kit-orange!30!black}
\colorlet{M4N16_color_multifuncrs}{kit-red}

\colorlet{M2N64_color}{kit-purple60}
\colorlet{M2N64_color_overlay}{kit-purple!40!black}
\colorlet{M2N64_color_multifuncrs}{kit-purple}

\colorlet{ebno2_hist}{kit-cyan}
\colorlet{ebno3_hist}{kit-green}
\colorlet{ebno4_hist}{kit-purple}

\pgfplotsset{
    colormap={coolsafe}{
        rgb255(0pt)=(45,106,79)
        rgb255(1pt)=(116,198,157)
        rgb255(2pt)=(180,230,190)
        rgb255(3pt)=(255,209,102)
        rgb255(4pt)=(239,140,42)
        rgb255(5pt)=(162,34,35)
        rgb255(6pt)=(120,20,20)
    }
}

\colorlet{light_gray_legend}{gray!80!black}
\colorlet{medium_gray_legend}{gray!30!black}
\colorlet{dark_gray_legend}{black}

\colorlet{31_16_color}{kit-cyan}
\colorlet{31_16_color_overlay}{kit-royalblue}
\colorlet{31_16_color_multifuncrs}{kit-blue}

\colorlet{15_7_color}{kit-lightgreen}
\colorlet{15_7_color_overlay}{kit-lightgreen!30!black}
\colorlet{15_7_color_multifuncrs}{kit-green}

\colorlet{7_4_color}{kit-orange}
\colorlet{7_4_color_overlay}{kit-orange!30!black}
\colorlet{7_4_color_multifuncrs}{kit-red}

\colorlet{63_30_color}{kit-purple60}
\colorlet{63_30_color_overlay}{kit-purple!40!black}
\colorlet{63_30_color_multifuncrs}{kit-purple}

\newtheorem*{remark}{Remark}

\tikzset{
    cube/.pic={
        \def\s{0.5}
        \def\dx{0.2}
        \def\dy{0.2}

        \fill[kit-orange!50] (0,\s) -- (\dx,\s+\dy) -- (\s+\dx,\s+\dy) -- (\s,\s) -- cycle;

        \fill[kit-orange!80] (\s,0) -- (\s+\dx,\dy) -- (\s+\dx,\s+\dy) -- (\s,\s) -- cycle;

        \fill[kit-orange!40] (0,0) rectangle (\s,\s);

        \draw[thick] (0,0) rectangle (\s,\s);
        \draw[thick] (\s,0) -- (\s+\dx,\dy);
        \draw[thick] (0,\s) -- (\dx,\s+\dy);
        \draw[thick] (\dx,\s+\dy) -- (\s+\dx,\s+\dy);
        \draw[thick] (\s+\dx,\dy) -- (\s+\dx,\s+\dy);
        \draw[thick] (\s,\s) -- (\s+\dx,\s+\dy);
    }
}

\tikzset{
    cube2/.pic={
        \def\s{0.5}
        \def\dx{0.2}
        \def\dy{0.2}

        \fill[kit-cyan!40] (0,\s) -- (\dx,\s+\dy) -- (\s+\dx,\s+\dy) -- (\s,\s) -- cycle;

        \fill[kit-cyan!60] (\s,0) -- (\s+\dx,\dy) -- (\s+\dx,\s+\dy) -- (\s,\s) -- cycle;

        \fill[kit-cyan!30] (0,0) rectangle (\s,\s);

        \draw[thick] (0,0) rectangle (\s,\s);
        \draw[thick] (\s,0) -- (\s+\dx,\dy);
        \draw[thick] (0,\s) -- (\dx,\s+\dy);
        \draw[thick] (\dx,\s+\dy) -- (\s+\dx,\s+\dy);
        \draw[thick] (\s+\dx,\dy) -- (\s+\dx,\s+\dy);
        \draw[thick] (\s,\s) -- (\s+\dx,\s+\dy);
    }
}

\usepackage{acronym}
\acrodef{AFDM}[AFDM]{affine frequency-division multiplexing}
\acrodef{APP}[APP]{a posteriori  probability}
\acrodefplural{APP}[APPs]{a posteriori  probabilities}
\acrodef{AWGN}[AWGN]{additive white Gaussian noise}
\acrodef{BCH}[BCH]{Bose–Chaudhuri–Hocquenghem}
\acrodef{BP}[BP]{belief propagation}
\acrodef{BPSK}[BPSK]{binary phase-shift keying}
\acrodef{BER}[BER]{bit error rate}
\acrodef{BMD}[BMD]{bit-metric decoder}
\acrodef{BMI}[BMI]{bitwise mutual information}
\acrodef{BICM}[BICM]{bit-interleaved coded modulation}
\acrodef{BI-AWGN}[BI-AWGN]{binary-AWGN}
\acrodef{BLER}[BLER]{block error rate}
\acrodef{CP}[CP]{canonical polyadic}
\acrodef{CSI}[CSI]{channel state information}
\acrodef{DMRG}[DMRG]{density matrix renormalization group}
\acrodef{DNN}[DNN]{deep neural network}
\acrodef{FEC}[FEC]{forward error correction}
\acrodef{FIR}[FIR]{finite impulse response}
\acrodef{EP}[EP]{expectation propagation}
\acrodef{iid}[i.i.d.]{independent and identically distributed}
\acrodef{ISI}[ISI]{inter-symbol interference}
\acrodef{LDPC}[LDCP]{low-density parity-check}
\acrodef{LLR}[LLR]{log-likelihood ratio}
\acrodef{LMMSE}[LMMSE]{linear minimum mean squared error}
\acrodef{MAP}[MAP]{maximum a posteriori}
\acrodef{MIMO}[MIMO]{multiple-input multiple-output}
\acrodef{ML}[ML]{maximum likelihood}
\acrodef{MLSE}[MLSE]{maximum likelihood sequence estimation}
\acrodef{MMSE}[LMMSE]{linear minimum mean squared error}
\acrodef{MPS}[MPS]{matrix product state}
\acrodef{MSE}[MSE]{mean squared error}
\acrodef{NN}[NN]{neural network}
\acrodef{NP}[NP]{nondeterministic polynomial-time}
\acrodef{OFDM}[OFDM]{orthogonal frequency-division multiplexing}
\acrodef{OSD}[OSD]{ordered statistics decoding}
\acrodef{pdf}[PDF]{probability density function}
\acrodef{pdp}[PDP]{power-delay profile}
\acrodef{pmf}[PMF]{probability mass function}
\acrodef{QAM}[QAM]{quadrature amplitude modulation}
\acrodef{QPSK}[QPSK]{quadrature phase-shift keying}
\acrodef{SER}[SER]{symbol error rate}
\acrodef{SNR}[SNR]{signal-to-noise ratio}
\acrodef{SPA}[SPA]{sum-product algorithm}
\acrodef{SVD}[SVD]{singular value decomposition}
\acrodef{TN}[TN]{tensor network}
\acrodef{TT}[TT]{tensor-train}
\acrodef{TTN}[TTN]{tree tensor network}
\acrodef{TTOpt}[TTOpt]{TT Optimizer}

\begin{document}
\title{A Tensor-Train Framework for Bayesian Inference in High-Dimensional Systems: Applications to MIMO Detection and Channel Decoding}

\author{
Luca Schmid, Dominik Sulz, Shrinivas Chimmalgi, and Laurent Schmalen,~\IEEEmembership{Fellow, IEEE}%
\thanks{The work of L. Schmid, S. Chimmalgi and L. Schmalen has received funding from the European Research Council (ERC) under the European Union’s Horizon 2020 research and innovation programme (grant agreement No. 101001899). The work of D. Sulz was funded by the Deutsche Forschungsgemeinschaft (DFG, German Research Foundation) – TRR 352 – Project-ID 470903074.}%
\thanks{L. Schmid and L. Schmalen are with the Communications Engineering Lab (CEL), Karlsruhe Institute of Technology (KIT), Hertzstr. 16, 76187 Karlsruhe, Germany (e-mail: \texttt{first.last@kit.edu}). C. Chimmalgi is now with Qoherent, Toronto, ON, Canada. D. Sulz is with the Technical University of Munich (TUM), Boltzmannstr. 3, 85748 Garching b. München, Germany (e-mail: \texttt{first.last@tum.de}).}%
}
\markboth{Submitted version, \today}%
{Submitted version, \today}
\maketitle

\begin{abstract} 
Bayesian inference in high-dimensional discrete-input additive noise models is a fundamental challenge in communication systems, as the support of the required joint a posteriori probability (APP) mass function grows exponentially with the number of unknown variables.
In this work, we propose a tensor-train~(TT) framework for tractable, near-optimal Bayesian inference in discrete-input additive noise models. The central insight is that the joint log-APP mass function admits an exact low-rank representation in the TT format, enabling compact storage and efficient computations.
To recover symbol-wise APP marginals, we develop a practical inference procedure that approximates the exponential of the log-posterior using a TT-cross algorithm initialized with a truncated Taylor-series. 
To demonstrate the generality of the approach, we derive explicit low-rank TT constructions for two canonical communication problems: the linear observation model under additive white Gaussian noise (AWGN), applied to multiple-input multiple-output (MIMO) detection, and soft-decision decoding of binary linear block error correcting codes over the binary-input AWGN channel. Numerical results show near-optimal error-rate performance across a wide range of signal-to-noise ratios while requiring only modest TT ranks.
These results highlight the potential of tensor-network methods for efficient Bayesian inference in communication systems.
\end{abstract}

\begin{IEEEkeywords}
    \noindent Tensor-trains, tensor networks, Bayesian inference, MIMO detection, channel decoding.
\end{IEEEkeywords}
\acresetall

\vspace{1cm}

\section{Introduction}\label{sec:intro}
Many problems in communications and signal processing reduce to probabilistic inference over noisy observations~\cite{proakis_digital_2007}. In most communication scenarios, the randomness of the channel, together with prior knowledge about the transmitter naturally leads to a Bayesian formulation at the receiver: the posterior distribution over the unknown quantities, e.g., the transmitted symbols, given the received signal is characterized and evaluated via its modes, marginals, or the model evidence~\cite{bishop_pattern_2006}. 

A fundamental challenge underlying Bayesian inference is the high dimensionality of the posterior distribution for large system sizes. For discrete-valued unknowns, the posterior can be interpreted as a multidimensional probability table whose order grows linearly with the number of variables while the number of entries grows exponentially. Explicit storage of this \ac{pmf} or any operation performed on it quickly becomes infeasible beyond small problem instances, a phenomenon known as the \emph{curse of dimensionality}. 

This challenge has motivated a substantial line of research on approximate inference algorithms, all of which have found application in communications.
Message passing algorithms such as \ac{BP}~\cite{kschischang_factor_2001,worthen_unified_2001,richardson_modern_2008} exploit the local structure of the \ac{pmf} to perform inference without explicitly computing the global distribution.
Variational methods and \ac{EP}~\cite{minka_expectation_2013,cespedes_expectation_2014} approximate the posterior within a restricted family of tractable distributions, such as Gaussian or factorized models.
Sampling-based techniques~\cite{doucet2005monte, farhang2006markov} avoid a closed-form posterior representation altogether and instead approximate expectations and marginals through Monte Carlo estimates.
More recently, deep learning has enabled flexible posterior approximations where intractable components of the distribution are parameterized by neural networks, as in variational autoencoders~\cite{kingma2013auto, lauinger_blind_2022} and diffusion models~\cite{ho2020denoising, fesl2024diffusion}.

A complementary perspective is to retain the posterior in its original form and instead seek a compact representation of this high-dimensional \ac{pmf}.
\Ac{TN} representations~\cite{Kressner_survey,hackbusch2012tensor} provide such a memory-efficient, data-sparse representation. By decomposing a high-order tensor into a network of connected low-order factors, \acp{TN} exploit the inherent low-rank structure of the underlying tensor and have proven effective across diverse fields, including supervised learning~\cite{stoudenmire2016supervised}, quantum many-body physics~\cite{haegeman2016unifying}, numerical optimization~\cite{holtz2012alternating}, and plasma physics~\cite{kormann2015semi}. Comprehensive surveys can be found in~\cite{Kressner_survey,kolda2009tensor}.

Early foundational works~\cite{cichocki2015tensor_SP,almeida2016overview,sidiropoulos2017tensor} recognized the paradigm shift towards high-dimensional signal representations via multiway arrays (i.e., tensors) and advocated tensor decompositions as a natural modeling tool for signal representations in modern communication systems.
Signals in modern communication and sensing systems are inherently multidimensional, including coupled domains such as time, frequency, space, and user indices. Tensor algebra enables these multi-way relationships to be represented and processed jointly, often leading to significantly more compact models than conventional approaches that treat dimensions separately~\cite{tokcan2025tensor}.

Motivated by this perspective, tensor methods have been applied to various communication problems, including channel estimation in \ac{MIMO}-\ac{OFDM}~\cite{araujo2019tensor} and \ac{MIMO}-\ac{AFDM} systems~\cite{wang2026tensor}, as well as parameter estimation in array processing and \ac{MIMO} radar~\cite{chen2021tensor}.
They have also been explored for user separation in unsourced massive random access, signal modulation design, and remote multidimensional sensing, as surveyed in~\cite{tokcan2025tensor}.
These works exploit the tensor structure in the signal or channel domain, representing received signals, propagation channels, or sensing data as low-rank tensors and leveraging this structure for efficient estimation or recovery. 

A conceptually distinct and, to our knowledge, largely unexplored direction is to apply tensor methods to the \emph{inference domain}. Instead of modeling the physical communication signals themselves, this perspective 
treats the posterior \ac{pmf} itself as the high-dimensional object to be represented and manipulated in a tensor format. Hence, it focuses the approximation directly on the probabilistic representation used for inference and decision making at the receiver, rather than on the underlying physical signal representation. 
This viewpoint naturally motivates the use of \ac{TN} representations to obtain compact and tractable approximations of the posterior distribution.

In this work, we investigate the potential of \acp{TN}, specifically \acp{TT}~\cite{oseledets2011tensor}, as a framework for efficient Bayesian inference in discrete-input additive noise models. Our central observation is that the \ac{APP} mass function of the transmitted symbols
often admits a low-rank representation in the \ac{TT} format when expressed in the logarithmic domain. Exploiting this structure enables computationally tractable inference in regimes where an explicit representation of the \ac{pmf} would otherwise be prohibitively large.

Specifically, we construct the log-posterior directly in the \ac{TT} format by building the log-likelihood contributions of all independent observations together with the log-prior using basic \ac{TT}~arithmetic. 
This construction in the logarithmic domain is exact with low \ac{TT} ranks. 
To obtain symbol-wise \acp{APP}, the log-posterior must subsequently be exponentiated and marginalized. 
While marginalization in the linear domain is straightforward in the \ac{TT} format, exponentiation is the main computational challenge. We address this step using a cross algorithm for \ac{TT}~\cite{Cross_survey}, initialized by a truncated Taylor-series.
This approximation is controllable: by increasing the maximum \ac{TT} rank, the resulting approximation can be made arbitrarily accurate, enabling near-optimal inference with complexity adaptable to the available computational resources.

We demonstrate the proposed framework on two well-known problems in communications. 
First, we consider \ac{MIMO} detection in the \ac{AWGN} channel. Second, we consider soft-decision decoding of binary linear block codes over the \ac{BI-AWGN} channel. For both problems, we derive explicit \ac{TT} constructions of the log-\ac{APP} and show that the resulting representations exhibit low ranks. 
Numerical results demonstrate near-optimal error-rate performance across wide \ac{SNR} regimes, in many cases outperforming established algorithms such as \ac{EP}-based detection and \ac{BP}-based decoding while requiring only modest \ac{TT} ranks.

We summarize the main contributions of this work:
\begin{itemize}
    \item We propose a \ac{TT} framework for Bayesian inference in discrete-input additive white noise models and provide an exact low-rank construction of the log-\ac{APP} mass function in the \ac{TT} format for two canonical models: the linear observation model under \ac{AWGN}, and channel decoding of binary linear block codes over the \ac{BI-AWGN} channel.
    \item We develop a practical method for marginal inference that recovers symbol-wise \acp{APP} by approximating the exponential of the log-posterior in \ac{TT} format via a truncated Taylor-series followed by a cross-approximation. We compare a variant of the \ac{TT}-cross algorithm and a \ac{DMRG}-like cross algorithm and analyze their trade-offs in terms of accuracy and robustness.
    \item We evaluate the proposed framework on \ac{MIMO} detection and soft-decision channel decoding, demonstrating near-optimal performance in terms of error rate across all \ac{SNR} regimes while requiring only modest \ac{TT} ranks.
\end{itemize}

The remainder of this paper is organized as follows. 
Section~\ref{sec:TT} introduces the \ac{TT} format and relevant arithmetics in the \ac{TT} format. Section~\ref{sec:main} describes the general discrete-input additive white noise observation model and the proposed \ac{TT} construction of the log-\ac{APP}, as well as the proposed inference method for exponentiation and marginalization. Sections~\ref{sec:mimo} and~\ref{sec:cc} specialize the framework to \ac{MIMO} detection and channel decoding, respectively, providing explicit \ac{TT} constructions and numerical evaluations. 
Section~\ref{sec:conclusion} concludes the paper.

\subsection*{Notation}
Throughout the paper, uppercase bold letters denote matrices $\bm{X}$ with entries $X_{ij}$ at row $i$ and column $j$.
Lowercase bold letters denote column vectors $\bm{x}$, whose $i$th element is written as $x_i$. 
$\lVert \cdot \rVert$ denotes the Euclidean norm, and~$d_\text{H}(\bm{x},\bm{y})$ denotes the Hamming distance between two vectors~$\bm{x},\bm{y}$.
The matrix transpose is denoted by~$(\cdot)^{\top}$ and~$\otimes$ is the Kronecker product for matrices.
The all-ones column vector of length $n$ is written as $\bm{1}_n$, and the ${n\times n}$ identity matrix is denoted by~$\bm{I}_n$.
For a vector~$\bm{x}$, the diagonal matrix with $\bm{x}$ on its main diagonal is denoted by~${\text{diag}\mleft(\bm{x}\mright)}$, e.g., ${\text{diag}\mleft(\bm{1}_n\mright) = \bm{I}_n}$.

We use uppercase nonbold letters to denote tensors~$A$. For consistency in this context, tensor slices, which may be matrices or vectors, are also denoted by nonbold symbols. In our notation, individual tensor entries are indexed using parentheses, e.g., $A(k_1, \dots, k_N)$.

We use ${\text{exp}\mleft(\cdot\mright)}$ to denote the exponential function, and ${\log \mleft(\cdot \mright)}$ and ${\log_{10} \mleft(\cdot \mright)}$ to denote the natural logarithm and the base-$10$ logarithm, respectively.
For a complex number ${c\in \mathbb{C}}$, ${\text{Re}\{ c \}}$ ${(\text{Im}\{ c \})}$ denotes its real (imaginary) part.
We use calligraphic letters to denote sets~$\mathcal{X}$ of cardinality $|\mathcal{X}|$. 
The indicator function~${\mathbbm{1}_{\{x = a\}}}$ equals $1$ if ${x=a}$ and $0$ otherwise.

The probability density function of a continuous random variable~${y}$ is denoted by $p_{{y}}(y)$ or $p(y)$ and the \ac{pmf} of a discrete random variable $x$ is $P_{{x}}(x)$ or $P(x)$. To keep the notation simple, we do not use a special notation for random variables since it is always clear from the context.
The Gaussian distribution, characterized by its mean $\mu$ and variance $\sigma^2$, 
is written as $\mathcal{N}(\mu,\sigma^2)$. The tail distribution function, or Q-function, of the standard normal distribution is denoted as~$Q(\cdot)$. The circular complex standard Gaussian distribution is denoted by~${\mathcal{CN}\mleft(0,1\mright)}$.
The noncentral chi-squared distribution with $k$ degrees of freedom and noncentrality parameter~$\lambda$ is denoted by~$\chi^2_k(\lambda)$. In case of ${\lambda=0}$, we simply write~$\chi^2_k$.

\section{The Tensor-Train Format}\label{sec:TT}
Tensors are a multidimensional generalization of matrices and are widely used in many applications~\cite{kolda2009tensor}. Due to their exponential scaling in memory requirements, tensors are typically approximated in a decomposed form. Classical tensor formats are the \ac{CP} decomposition \cite{hitchcock1927_CP}, Tucker tensors \cite{tucker1966some}, \acfp{TT} \cite{oseledets2011tensor} and the more general class of \acp{TTN}~\cite{hackbusch2012tensor}. 
In this work, we consider \acp{TT}, which are also known as \acp{MPS} in other fields~\cite{MPS_paper}. 

\begin{figure}[tb]
\centering
\begin{tikzpicture}[
    transform shape,
    core/.style={circle, draw=black, fill=orange!40, thick, minimum size=1.8em, inner sep=0pt},
    link/.style={very thick},
    phys/.style={very thick},
]

\node[core] (G1) at (0,0) {$G_{1}$};
\node[core, right=0.75cm of G1] (G2) {$G_{2}$};
\node[core, right=0.75cm of G2] (G3) {$G_3$};
\node[core, right=0.75cm of G3] (G4) {$G_4$};
\node[core, right=1.5cm of G4] (G5) {$G_N$};

\draw[link] (G1) -- (G2)node[midway, above]{$r_1$};
\draw[link] (G2) -- (G3)node[midway, above]{$r_2$};
\draw[link] (G3) -- (G4)node[midway, above]{$r_3$};
\draw[link] (G4) -- ++(0.75cm,0) node[midway, above]{$r_4$};
\draw[link, dotted] (G4) -- (G5)node[midway, above]{};
\draw[link] (G5) -- ++(-0.75cm,0) node[pos=-0.4, above, anchor=south east]{$r_{N-1}$};

\draw[link] (G1.west) -- ++(-0.75,0) node[midway, above]{$r_0$};
\draw[link] (G5.east) -- ++(0.75,0)node[midway, above]{$r_N$};

\draw[phys] (G1.south) -- ++(0,-0.4) node[below] {$n_1$};
\draw[phys] (G2.south) -- ++(0,-0.4) node[below] {$n_2$};
\draw[phys] (G3.south) -- ++(0,-0.4) node[below] {$n_3$};
\draw[phys] (G4.south) -- ++(0,-0.4) node[below] {$n_4$};
\draw[phys] (G5.south) -- ++(0,-0.4) node[below] {$n_N$};

\end{tikzpicture}
\caption{Graphical representation of the \ac{TT} decomposition. Each edge can be interpreted as a tensor contraction.}
\label{fig:TT_representation}
\end{figure}
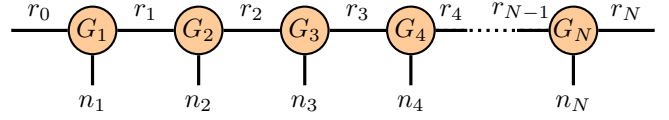

Consider a tensor ${A \in \R^{n_1 \times \dots \times n_N}}$ of order~$N$ with the physical dimensions $n_i, \; i=1,\ldots, N$. The memory requirement of the full tensor~$A$ in explicit form scales exponentially with~$N$.
The decomposition of~$A$ in the \ac{TT} format is given by
\[
A(k_1,\dots,k_N) = G_1(k_1) G_2(k_2)\cdots G_N(k_N),
\]
with~${G_i(k_i) \in \R^{r_{i-1} \times r_i}}$, ${i=1,\dots,N}$, and ${r_0 = r_N = 1}$. Equivalently, the \ac{TT} decomposition can be expressed in index notation as
\begin{align*}
&A(k_1,\dots,k_N) \\ &= \; \sum_{\mathclap{j_0,j_1,\dots,j_N}} \; G_1(j_0,k_1,j_1) G_2(j_1,k_2,j_2) \cdots G_N(j_{N-1},k_N,j_N).
\end{align*}
Fig.~\ref{fig:TT_representation} visualizes the \ac{TT} decomposition of a tensor~$A$ into a sequence of core tensors~$G_i$. 
For indexing the core tensors~${G_i \in \R^{r_{i-1} \times n_i \times r_i}}$ of order~$3$, 
we follow the standard nomenclature in \ac{TN} literature~\cite{oseledets2011tensor}: The second dimension representing the physical index~$n_i$, is oriented along the $z$-axis (depth). The first and third dimensions correspond to the auxiliary rank indices~$r_{i-1}$ and~$r_i$, which encode the connection between adjacent cores, respectively, as illustrated in Fig.~\ref{fig:TT_representation}.
The matrices $G_i(k_j) =: G_i(:,k_j,:)$ for $k_j=1,\ldots,n_i$ are often called \textit{slices} of the tensor $G_i$. 

The maximum rank of a \ac{TT} is defined as ${r_{\text{max}} = \max_{i=0,\ldots N} r_i}$. 
While any tensor theoretically admits a \ac{TT} representation, the format only yields significant gains in storage and computational efficiency when the ranks of the underlying tensor, or an approximation thereof, remain small. For a more intuitive grasp of the \ac{TT} format, we refer the reader to Appendix~\ref{appendix:ex}, where we provide an illustrative example.

\subsection{Arithmetics in the Tensor-Train Format}\label{subsec:TT_computations}
Many basic operations, such as addition and element-wise multiplication, can directly be performed in the \ac{TT} format, without forming the full tensor. 
We briefly summarize the most basic operations, which can be originally found in \cite{oseledets2011tensor}. 

Consider two \acp{TT}~${A = A_1(k_1) \cdots A_N(k_N)}$ and ${B = B_1(k_1) \cdots B_N(k_N)}$ and denote by ${(r_i^A)_{i=0}^N}$ and ${(r_i^B)_{i=0}^N}$ their respective ranks. 

\subsubsection{\textbf{Addition}} The sum ${C = A + B}$ can be computed as
\begin{align*}
    C_i(k_i) &= \begin{pmatrix}
        A_i(k_i) & 0 \\ 0 & B_i(k_i)
    \end{pmatrix}, \ &i=2,\dots,N-1, \\
    C_1(k_1) & = \begin{pmatrix}
        A_1(k_1) & B_1(k_1)
    \end{pmatrix}, &C_N(k_N) = \begin{pmatrix}
        A_N(k_N) \\ B_N(k_N)
    \end{pmatrix}.
\end{align*}
This strategy requires no arithmetic operations but increases the ranks of $C$ to ${r_i^C = r_i^A + r_i^B,}$ ${i=1,\dots,N-1}$. Typically, a rank truncation is performed after the addition, which finds a rank-reduced approximation, see Sec.~\ref{subsec:truncation}. 

\subsubsection{\textbf{Tensor-matrix multiplication}}
Let ${A \in \R^{n_1 \times \dots \times n_N}}$ be a full tensor and $\bm{U} \in \R^{m \times n_i}$ a matrix. The tensor-matrix multiplication $A \times_i \bm{U}$ is defined by~\cite{LMV2000_HOSVD}
\begin{align*}
    &(A \times_i \bm{U})(k_1,\dots,k_i,\dots,k_N) \\ &:= \sum_{j=1}^{n_i} A(k_1,\dots,k_{i-1},j,k_{i+1},\dots,k_N) (\bm{U}^\top \! )_{j k_i},
\end{align*}
for $k_i = 1,\dots,m$. This operation can be interpreted as the contraction of the matrix $\bm{U}$ with the tensor $A$ in the $i$th mode. In the following, we slightly abuse the notation $\times_i$ to denote the tensor-matrix multiplication in the \ac{TT} format. 
Specifically, ${A \times_i \bm{U}}$ is defined as the \ac{TT} obtained from $A$ by replacing its $i$th core $G_i$ with the tensor ${G_i \times_2 \bm{U}}$.

\subsubsection{\textbf{Marginalization}} \label{subsubsec:marginalization}
The marginalization in the $i$th component of a tensor $A$ is the summation over all dimensions except $i$, which is mathematically equivalent to 
\begin{align*}
    A \bigtimes_{\substack{j=1\\ j \neq i}}^N \mathbf{1}_{n_j}^\top, \qquad \mathbf{1}_{n_j} = (1,\dots,1)^\top \in \R^{n_j},
\end{align*}
where
\begin{align*}
    A \bigtimes_{j=1}^N \bm{U}_j = A \times_1 \bm{U}_1 \dots \times_N \bm{U}_N.
\end{align*}

\subsubsection{\textbf{Element-wise multiplication}}
The element-wise multiplication or Hadamard product ${C = A \circ B}$ is computed by 
\begin{align*}
    C_i(k_i) = A_i(k_i) \otimes B_i(k_i), \qquad i=1,\dots,N.
\end{align*}
The Kronecker product for matrices~$\otimes$ causes a multiplication of the ranks of the two tensors $A$ and $B$ in the \ac{TT} format, i.e., ${r_i^C = r_i^A \cdot  r_i^B}$ for $i=0,\dots,N$.

\subsubsection{\textbf{Scalar multiplication}} 
The product~${\lambda A}$ of a tensor~$A$ in the \ac{TT} format with a scalar~${\lambda \in \R}$ is obtained by multiplying $\lambda$ to one of the core tensors of~$A$. It does not matter to which of the $N$ core tensors the multiplication is applied, and the rank is not increased.

\subsubsection{\textbf{Truncation}} \label{subsec:truncation}
Many operations, such as addition or element-wise multiplication of tensors in the \ac{TT} format, significantly increase the ranks. To keep computations and memory feasible, these operations are typically followed by a rank truncation. Common truncation algorithms use consecutive singular value decompositions of the core tensors. By discarding singular values smaller than a tolerance $\vartheta>0$, the tensor is typically compressed to a \ac{TT} representation of lower ranks. This procedure ensures that the approximation error remains bounded and is directly controlled by the choice of $\vartheta$, as the error grows linearly with $\vartheta$ \cite{hackbusch2012tensor,oseledets2011tensor}. Since rank truncation is a standard operation in the context of tensor networks, we refer to \cite{oseledets2011tensor} for a detailed description.

\subsubsection{\textbf{TT-cross algorithm}} \label{subsubsec:ttcross}
Another key component of the proposed method is the element-wise application of a nonlinear function ${f(A)}$, where $A$ is a tensor in \ac{TT} format. 
We consider \ac{TT}-cross algorithms~\cite{oseledets_TT_cross,ghahremani2024deim} to evaluate $f(A)$ directly in the \ac{TT} format.
The \ac{TT}-cross can be interpreted as an interpolation scheme for multivariate functions on a tensor grid. This method allows evaluations without forming the full tensor. 
Since a detailed description of the algorithm would obscure the main ideas of this work, we refer to Appendix~\ref{app:cross} and \cite{oseledets_TT_cross} for a more detailed discussion and to~\cite{Cross_survey} for a broader survey of tensor cross algorithms.

\begin{remark}
Our considerations can be extended to the more general class of tree tensor networks (TTN), which were first introduced in quantum chemistry \cite{wang2003multilayer}. In mathematical literature, they were considered first as hierarchical Tucker tensors in the binary case 
\cite{hackbusch2012tensor,hackbusch_first_HT_paper} and in \cite{falco2015geometric} for the general case. We note that both binary and general TTNs include \acp{TT} as a special case. While TTNs often allow representations with lower ranks, their implementation is considerably more involved.
\end{remark}

\section{Detection in Additive Noise Models}\label{sec:main}
\subsection{System Model}\label{subsec:additive_noise_channel_model}
We consider the real-valued\footnote{All considerations can trivially be extended to complex-valued models. For the sake of exposition, we only consider the real-valued case.}
observation model with additive white noise
\begin{equation}
    \bm{y} = f(\bm{x}) + \bm{n}, \qquad \bm{x} \in \mathcal{A}^{\NT}, \, \bm{n} \in \R^{\NR}
\label{eq:linear_model}
\end{equation}
where~$n_1,\ldots,n_{\NR}$ are \ac{iid} noise samples from an exponential family distribution.
The transmit sequence~$\bm{x}$ consists of \ac{iid} symbols $x_i$, ${i=1,\ldots,\NT}$, each drawn from a discrete alphabet~${\mathcal{A} \subset \R}$ of cardinality ${|\mathcal{A}|=\Mike}$ with probability~${P(x_i)}$.
In the context of Bayesian inference, we are interested in the symbol-wise \acp{APP}
\begin{align}
    P(x_i = a | \bm{y}) = \sum\limits_{\mathclap{\substack{\bm{a} \in \mathcal{A}^{\NT}\\ a_i = a}}} P(\bm{x}=\bm{a} | \bm{y}), \; a\in \mathcal{A}, \, i = 1,\ldots, \NT. \!
    \label{eq:marginalization}
\end{align}
The symbol-wise \ac{MAP} detector, which minimizes the symbol error probability, is defined as
\begin{align}
\hat{x}_{i,\text{MAP}} := \arg\max\limits_{a \in \mathcal{A}} P(x_i=a | \bm{y}).
\label{eq:max}
\end{align}
Using Bayes' theorem, the \ac{APP} distribution can be written as
\begin{align*}
    P(\bm{x}|\bm{y}) \propto p(\bm{y}|\bm{x}) P(\bm{x}) 
    \propto 
    \prod\limits_{j=1}^{\NR} \mathrm{e}^{ \ell(y_j|\bm{x})}
    \prod\limits_{i=1}^{\NT} P(x_i)
\end{align*}
where ${\ell(y_j|\bm{x})}$ denotes the log-likelihood of the observation~$y_j$, and the proportionality~$\propto$ means that two terms differ only in a factor independent of~$\bm{x}$.
Equivalently, the \ac{APP} can be expressed in the logarithmic domain as
\begin{equation}
    \log P(\bm{x}|\bm{y}) = C + \sum\limits_{j=1}^{\NR} \ell(y_j|\bm{x})
    + \sum\limits_{i=1}^{\NT} \log P(x_i) =: C + \Lambda(\bm{x}).
    \label{eq:log_app}
\end{equation}
Since~$C$ is a constant with respect to~$\bm{x}$, we can neglect it in the context of detection and instead consider the unnormalized log-\ac{APP} metric~${\Lambda(\bm{x})}$ in the following.

\subsection{Main Contribution} \label{subsec:main_contribution}
Our proposed approach represents the log-likelihood terms~$\ell(y_j|\bm{x})$ and the log-prior~$\log P(\bm{x})$ exactly in the \ac{TT} format with low ranks.
In this framework, the tensor structure serves as a probability lookup table, where each entry of the tensor in $\R^{L \times \dots \times L}$ corresponds to an evaluation of the discrete \ac{APP} mass function.
If the problem admits a low-rank \ac{TT} decomposition or approximation, this representation enables efficient Bayesian inference in high-dimensional settings.
We construct the \ac{APP} in the logarithmic domain according to~\eqref{eq:log_app}, as this typically yields much lower \ac{TT} ranks compared to the linear domain.
The symbol-wise \acp{APP} are then obtained by exponentiation and marginalization, with all operations carried out directly in the \ac{TT} format, as described in Sec.~\ref{subsec:TT_computations}.

\subsubsection{\textbf{Construction of ${\log P(\bm{x}) = \sum_{i=1}^{\NT} \log P(x_i)}$}} \label{subsubsec:priors}
The prior can be exactly represented by a rank-$2$ \ac{TT}, where the rank is independent of $\NR$ and $\NT$. We denote the vector of log-priors for any symbol~$x_i$ by $\bm{v} = \left(\log P(x_i=a_1),\ldots, \log P(x_i=a_L)\right)^\top \in \R^{L}$, where $\{a_1,\ldots,a_L\}=\mathcal{A}$. The \ac{TT} construction has the first and last core
\begin{align*}
    G_1 &= \begin{pmatrix}
         \mathbf{1}_L & \bm{v}
    \end{pmatrix} \in \R^{1 \times L \times 2}, \\
    G_{\NT} &= \begin{pmatrix}
        \bm{v}^\top \\
        \mathbf{1}_L^\top
    \end{pmatrix} \in \R^{2 \times L \times 1}.
\end{align*}
The remaining cores $G_i \in \R^{2\times L \times 2}$, for $i=2,\dots,\NT-1$, can be constructed by 
\begin{align*}
    G_i(:,j,:) = \begin{pmatrix}
        1 & v_j \\ 0 & 1
    \end{pmatrix}, \quad \text{for} \ j=1,\dots,L.
\end{align*}

\subsubsection{\textbf{Sum over log-likelihoods}} \label{subsubsec:sum_over_log_likelihoods}
For a low-rank \ac{TT} representation of the sum over the log-likelihood terms in~\eqref{eq:log_app}, we consider two strategies: constructing the full sum directly in the \ac{TT} format, or constructing each log-likelihood term $\ell(y_j|\bm{x})$, $j=1,\dots,\NR$, separately and summing the resulting $\NR$ \ac{TT}s. 
Both strategies are mathematically equivalent; the choice is problem-dependent and can be based on the simplicity of the construction, respectively. 

If each log-likelihood term~$\ell(y_j|\bm{x})$, ${i=1,\ldots,\NR,}$ is constructed individually in the \ac{TT} format of rank at most~${r}$,
summing the $N_\text{r}$ log-likelihood terms also sums the ranks, yielding a result of rank at most~$r \NR$. 
The rank of the directly constructed full sum typically scales with $\NR$ as well. 
In either case, a truncation with tolerance~$\vartheta>0$ may recompress the resulting tensor to a moderate rank. 
Note that the above rank estimates are only upper bounds. %

Finally, we add the log-prior~$\log P(\bm{x})$ to the joint log-likelihood, completing the \ac{TT} construction of the full log-\ac{APP} in \eqref{eq:log_app}.

\subsubsection{\textbf{Exponentiation}} \label{subsubsec:ttexp}
Let $A$ denote the \ac{TT} representation of the log-\ac{APP} metric~${\Lambda(\bm{x})}$ in~\eqref{eq:log_app}, constructed as described above. 
Our goal is to compute~$\exp(A)$, however, since $\exp(\cdot)$ is nonlinear, the ranks of an exact representation of $\exp(A)$ can become prohibitively large. 
Therefore, we approximate $\exp(A)$ using a \ac{TT}-cross approach (cf. Sec.~\ref{subsubsec:ttcross}). We  consider two \ac{TT}-cross algorithms, a classical \ac{TT}-cross variant and a \ac{DMRG}-like cross algorithm. For the numerical evaluation we use the implementations~\texttt{multifuncrs} and \texttt{funcrs} from the \ac{TT}-toolbox~\cite{oseledets2016_toolbox}, respectively. 

To improve the quality of the approximation, both methods support an initialization step for the \ac{TT}-cross algorithm. 
We use a truncated Taylor series
\[
    \exp(A) \approx \sum\limits_{k=0}^{p} \frac{A^k}{k!},
\]
for initialization, where $A^k$ denotes the element-wise power. For this, we use the~\texttt{tt\_exp} implementation from the \ac{TT}-toolbox~\cite{oseledets2016_toolbox}.
Typically, $p\approx 10$ provides a reliable initialization. The \texttt{tt\_exp} function also accepts a maximum rank parameter \texttt{r\_{max}} to keep the computation tractable.

The two cross functions differ in their underlying algorithmic strategy. 
The \texttt{multifuncrs} routine follows a more classical \ac{TT}-cross interpolation approach, reconstructing the tensor from adaptively sampled entries. In contrast, \texttt{funcrs} employs a \ac{DMRG}-like sweeping algorithm that updates local \ac{TT} cores through low-rank projections, while still falling into the class of cross approximations. 
The latter typically achieves higher accuracy at the cost of increased computational complexity. A central ingredient in both cross functions is the use of random sampling in each core. This enables the addition of new directions to avoid getting stuck in local minima. Consequently, both functions are non-deterministic.

\subsubsection{\textbf{Marginalization}}\label{subsubsec:marginalization}
To obtain the symbol-wise \acp{APP} in~\eqref{eq:marginalization}, the marginalization over all dimensions except for one needs to be computed.
This can efficiently be done in the \ac{TT} format as described in Sec.~\ref{subsubsec:marginalization}.
The result of the marginalization is a tensor of dimension 1, i.e., a vector that represents the entries proportional to~${P(x_i=a | \bm{y})}$, $a\in \mathcal{A}$.

\subsection{Computational Complexity}
The computational complexity is dominated by the \ac{TT}-cross algorithm and the truncation function, both of which rely on successive \acp{SVD} of the underlying core tensors.
If the core tensors are of size $r \times n \times r$, the computational complexity of a single \ac{SVD} scales with~$\mathcal{O}(r^3)$, where $r \leq r_{\text{max}}$. One full sweep through the \ac{TT} scales with $\mathcal{O}(Nr^3)$. Such sweeps are performed iteratively, for instance, to evaluate the terms of the truncated Taylor series.

\begin{remark}
    Instead of computing the exponential of $A$ followed by a marginalization, an alternative approach is to find the maximal element of the full tensor~$A$ in the log-domain, which corresponds to \ac{MAP} sequence estimation. 
    A promising candidate for solving this maximization problem is the \ac{TTOpt}~\cite{sozykin2022ttopt}, which provides a gradient-free and efficient optimization approach in the \ac{TT} format.
    We leave this maximization-based approach for future investigation and focus here on the marginalization, i.e., the symbol-wise detection problem.
\end{remark}

In the following, we consider two inference problems relevant to the field of communications: \ac{MIMO} detection and channel decoding for \ac{AWGN} channels. We provide explicit log-\ac{APP} constructions in the \ac{TT} format and thereby demonstrate that these problems inherently admit low-rank representations.

\section{Example: MIMO Detection}\label{sec:mimo}
\Ac{MIMO} systems are a key technology in many current and emerging wireless communication systems due to their high spectral efficiency and throughput~\cite{yang_fifty_2015}.
In this context, efficient symbol detection is a crucial computational bottleneck in realizing these theoretical performance gains in practical systems~\cite{cespedes_expectation_2014}.

\subsection{MIMO Channel Model}\label{subsec:mimo_channel_model}
We consider a \ac{MIMO} system with $\NTc$ antennas at the transmitter and $\NRc$~antennas at the receiver. 
The transmit symbols~${\tilde{x}_i}$, ${i=1,\ldots,\NTc}$ are independently and uniformly drawn from an $M$-\ac{QAM} constellation~${\mathcal{M} \subset \mathbb{C}}$.
The channel output~${\tilde{\bm{y}} \in \mathbb{C}^{\NRc}}$ is given by ${\tilde{\bm{y}} = \tilde{\bm{H}} \tilde{\bm{x}} + \tilde{\bm{n}}}$, where $\tilde{\bm{n}}$ is circular complex \ac{AWGN}.
We assume a Rayleigh-fading \ac{MIMO} channel, i.e., each element of~$\tilde{\bm{H}}$ is independently sampled from a circular complex standard Gaussian distribution~$\mathcal{CN}(0,1)$.
We define the receiver \ac{SNR} of each transmission as
\begin{equation*}
    \text{SNR} := 10 \log_{10} \mleft( \frac{ \| \tilde{\bm{H}} \tilde{\bm{x}} \|^2} {\| \tilde{\bm{n}} \|^2} \mright) \quad \text{in dB}.
\end{equation*}

The complex-valued \ac{MIMO} system can be decomposed into an equivalent real-valued representation 
\begin{equation*}
    \bm{H} = 
     \begin{pmatrix} \text{Re}\{\tilde{\bm{H}}\} & - \text{Im}\{\tilde{\bm{H}}\} \\ \text{Im}\{\tilde{\bm{H}}\} & \text{Re}\{\tilde{\bm{H}}\} \end{pmatrix}
     \in \mathbb{R}^{\NR \times \NT}, 
     \quad 
     \begin{matrix}
\NT = 2 \NTc,\\
\NR = 2 \NRc,
\end{matrix}
\end{equation*}
to match the model in~\eqref{eq:linear_model} with
$\bm{y} = (\text{Re}\{\tilde{\bm{y}}\}, \text{Im}\{\tilde{\bm{y}}\})^\top$, $\bm{x} = (\text{Re}\{\tilde{\bm{x}}\}, \text{Im}\{\tilde{\bm{x}}\})^\top$, $f(\bm{x})=\bm{H}\bm{x}$, $\bm{n} = (\text{Re}\{\tilde{\bm{n}}\}, \text{Im}\{\tilde{\bm{n}}\})^\top$, and
$\mathcal{A} = \{\pm 1, \pm 3, \ldots \}$ with $L=|\mathcal{A}| = \sqrt{M}$.

In general, \Ac{MIMO} detection refers to the task of inferring the transmit vector~$\bm{x}$ from the channel observation~$\bm{y}$~\cite{yang_fifty_2015}, and is a proven \ac{NP}-hard problem~\cite{verdu_computational_1989}. 
In this work, we assume perfect \ac{CSI} at the receiver, i.e., knowledge of $\bm{H}$ and the \ac{SNR}.

In the following, we consider \emph{symbol-wise} \ac{MAP} \ac{MIMO} detection, which minimizes the symbol error probability, and requires the computation of the marginal \acp{APP} ${P(x_i|\bm{y})}$, cf.~\eqref{eq:marginalization}. This marginalization involves the summation over an exponentially large set of transmit vectors~$\bm{x}$ and is therefore \#P-hard in general.

\begin{remark}
    The following considerations regarding symbol-wise \ac{MAP} detection for \ac{MIMO} channels apply to any real-valued linear observation model with \ac{AWGN}
\begin{equation}
\bm{y} = \bm{H} \bm{x} + \bm{n}, \qquad \bm{n} \sim \mathcal{N}(\bm{0}, \sigma^2 \bm{I}_{\NR}),
\label{eq:MIMO_model}
\end{equation}
where~$\bm{y}$ is the received signal, $\bm{H} =: (\bm{h}_1,\ldots,\bm{h}_{\NR})^\top \in \R^{\NR \times \NT}$~is the observation matrix, and $\bm{x}$ is the discrete-valued vector of unknowns.
\end{remark}

\subsection{TT Construction of the Log-Likelihood Terms} \label{subsec:mimo_log_likelihood_construction}
Our objective is to express the unnormalized log-\ac{APP} metric~${\Lambda(\bm{x})}$ in~\eqref{eq:log_app} in the \ac{TT} format.
The construction of the log-priors ${\log P(\bm{x})}$, which is described in Sec.~\ref{subsubsec:priors}, can be omitted in case of uniform priors, as assumed here. This section focuses on the specific construction of the log-likelihood terms~$\ell(y_j|\bm{x})=-(y_j - \mathbf{h}_j^\top \bm{x})^2/(2\sigma^2)$, ${j=1,\ldots,\NR}$ for the linear observation model~\eqref{eq:MIMO_model}, such as the \ac{MIMO} channel model.

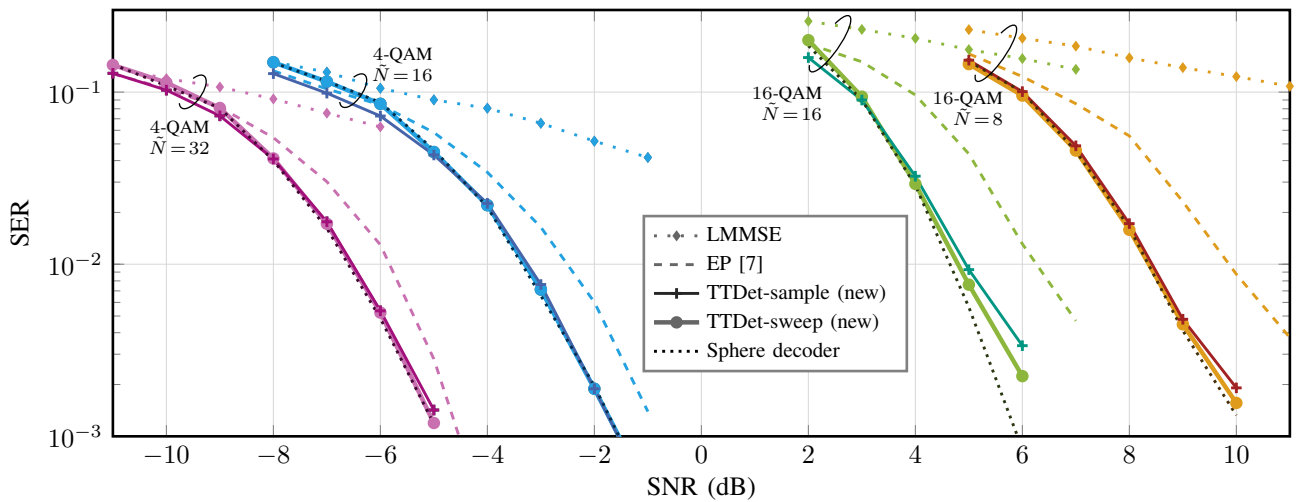
\begin{figure*}[t]
\centering
  \begin{tikzpicture}
  \begin{axis}[
    width=0.95\linewidth, %
    height=0.4\linewidth,
    align = left,
    grid=major, %
    grid style={gray!30}, %
    xlabel= SNR (dB),
    xlabel style={yshift=3pt, xshift=0pt},
    ylabel= SER,
    ymode=log,
    ymax = 0.3,
    ymin = 0.001,
    xmin = -11,
    xmax = 11,
    enlarge x limits=false,
    enlarge y limits=false,
    line width=1pt,
	  legend style={font=\footnotesize, cells={align=left}, draw=gray, anchor=south west, at={(0.45,0.15)}},
    legend cell align={left},
    mark options={solid, mark size=1.1pt},
    every axis plot/.append style={line width=1.1pt },
    axis line style={line width=1pt},
    axis background/.style={fill=white},
    ]
     \addplot[draw=none, light_gray_legend, loosely dotted, mark=diamond*,mark options={solid, mark size=1.2pt}] table[x={SNR (dB)}, y={SER LMMSE}, col sep=comma] {numerical_results/M2_N32_results_EP.csv};
    \addplot[draw=none, light_gray_legend, dashed, mark=none] table[x={SNR (dB)}, y={SER EP}, col sep=comma] {numerical_results/M2_N32_results_EP.csv};
    \addplot[draw=none,medium_gray_legend, mark=+, mark options={mark size=2pt}] table[x={SNR (dB)}, y={SER}, col sep=comma] {numerical_results/M2_N64_rmax10_results_tt_multifuncrs.csv};
    \addplot[draw=none, light_gray_legend, mark=*, mark options={mark size=1.5pt}, line width=1.7pt] table[x={SNR (dB)}, y={SER}, col sep=comma] {numerical_results/M2_N32_rmax10_results_tt.csv};
    \addplot[draw=none, dark_gray_legend, dotted] table[x={SNR (dB)}, y={SER}, col sep=comma] {numerical_results/M2_N32_results_sphere_decoder.csv};
    \legend{LMMSE \\ EP~\cite{cespedes_expectation_2014} \\ TTDet-sample (new) \\ TTDet-sweep (new) \\ Sphere decoder \\}
    \addplot[M2N32_color, loosely dotted, mark=diamond*,mark options={solid, mark size=1.2pt}] table[x={SNR (dB)}, y={SER LMMSE}, col sep=comma] {numerical_results/M2_N32_results_EP.csv};
    \addplot[M2N32_color, dashed, mark=none] table[x={SNR (dB)}, y={SER EP}, col sep=comma] {numerical_results/M2_N32_results_EP.csv};
    \addplot[M2N32_color, mark=*, mark options={mark size=1.5pt}, line width=1.7pt] table[x={SNR (dB)}, y={SER}, col sep=comma] {numerical_results/M2_N32_rmax10_results_tt.csv};
    \addplot[M2N32_color_multifuncrs, mark=+, mark options={mark size=2pt}] table[x={SNR (dB)}, y={SER}, col sep=comma] {numerical_results/M2_N32_rmax10_results_tt_multifuncrs.csv};
     \addplot[M2N32_color_overlay, dotted] table[x={SNR (dB)}, y={SER}, col sep=comma] {numerical_results/M2_N32_results_sphere_decoder.csv};
    \draw[line width=0.5pt, rotate around={55:(axis cs:-6.5,0.122)}] (axis cs: -6.5, 0.122) arc(55:-210:0.26cm and 0.12cm) node[above right =5mm and 3mm, anchor=west, font=\scriptsize, align=right]{$4$-QAM\\${\tilde{N}\!=\!16}$};
    \addplot[M2N64_color, loosely dotted, mark=diamond*, restrict x to domain=-inf:-6,mark options={solid, mark size=1.2pt}] table[x={SNR (dB)}, y={SER LMMSE}, col sep=comma] {numerical_results/M2_N64_results_EP.csv};
     \addplot[M2N64_color, dashed, mark=none] table[x={SNR (dB)}, y={SER EP}, col sep=comma] {numerical_results/M2_N64_results_EP.csv};
     \addplot[M2N64_color, mark=*, mark options={mark size=1.5pt}, line width=1.7pt] table[x={SNR (dB)}, y={SER}, col sep=comma] {numerical_results/M2_N64_rmax10_results_tt.csv};
     \addplot[M2N64_color_multifuncrs, mark=+, mark options={mark size=2pt}] table[x={SNR (dB)}, y={SER}, col sep=comma] {numerical_results/M2_N64_rmax10_results_tt_multifuncrs.csv};
    \addplot[M2N64_color_overlay, dotted,restrict x to domain=-11:-5] table[x={SNR (dB)}, y={SER}, col sep=comma] {numerical_results/M2_N64_results_sphere_decoder.csv};
    \draw[line width=0.5pt, rotate around={55:(axis cs:-9.5,0.12)}] (axis cs:-9.5,0.12) arc(55:-210:0.26cm and 0.12cm) node[below =1.0mm, anchor=north, font=\scriptsize , align=right]{$4$-QAM\\${\tilde{N}\!=\!32}$};
    \addplot[M4N16_color, loosely dotted, mark=diamond*,mark options={solid, mark size=1.2pt}] table[x={SNR (dB)}, y={SER LMMSE}, col sep=comma] {numerical_results/M4_N16_results_EP.csv};
     \addplot[M4N16_color, dashed, mark=none] table[x={SNR (dB)}, y={SER EP}, col sep=comma] {numerical_results/M4_N16_results_EP.csv};
     \addplot[M4N16_color, mark=*, mark options={mark size=1.5pt}, line width=1.7pt] table[x={SNR (dB)}, y={SER}, col sep=comma] {numerical_results/M4_N16_rmax30_results_tt.csv};
     \addplot[M4N16_color_multifuncrs, mark=+, mark options={mark size=2pt}] table[x={SNR (dB)}, y={SER}, col sep=comma] {numerical_results/M4_N16_rmax20_results_tt_multifuncrs.csv};
    \addplot[M4N16_color_overlay, dotted] table[x={SNR (dB)}, y={SER}, col sep=comma] {numerical_results/M4_N16_results_sphere_decoder.csv};
    \draw[line width=0.5pt, rotate around={55:(axis cs:5.6,0.22)}] (axis cs:5.6,0.22) arc(55:-210:0.45cm and 0.12cm) node[below left=1.0mm and 1mm, anchor=north, font=\scriptsize,  align=right]{$16$-QAM\\${\tilde{N}\!=\!8}$};
    \addplot[M4N32_color, loosely dotted, mark=diamond*,mark options={solid, mark size=1.2pt}] table[x={SNR (dB)}, y={SER LMMSE}, col sep=comma] {numerical_results/M4_N32_results_EP.csv};
     \addplot[M4N32_color, dashed, mark=none] table[x={SNR (dB)}, y={SER EP}, col sep=comma] {numerical_results/M4_N32_results_EP.csv};
     \addplot[M4N32_color, mark=*, mark options={mark size=1.5pt}, line width=1.7pt] table[x={SNR (dB)}, y={SER}, col sep=comma] {numerical_results/M4_N32_rmax40_results_tt.csv};
     \addplot[M4N32_color_multifuncrs, mark=+, mark options={mark size=2pt}] table[x={SNR (dB)}, y={SER}, col sep=comma] {numerical_results/M4_N32_rmax40_results_tt_multifuncrs.csv};
    \addplot[M4N32_color_overlay, dotted] table[x={SNR (dB)}, y={SER}, col sep=comma] {numerical_results/M4_N32_results_sphere_decoder.csv};
    \draw[line width=0.5pt, rotate around={55:(axis cs:2.5,0.25)}] (axis cs:2.5,0.25) arc(55:-210:0.45cm and 0.12cm) node[below left =1.5mm and 3mm, anchor=north, font=\scriptsize , align=right]{$16$-QAM\\${\tilde{N}\!=\!16}$};
    \end{axis}
\end{tikzpicture}
    \caption{\Ac{SER} over \ac{SNR} for various \ac{MIMO} detectors across different ${\tilde{N}\times \tilde{N}}$ system configurations (real-valued dimensions ${\NT=\NR=2\tilde{N}}$). The maximum rank~\texttt{r\_max} in the truncated Taylor-series initialization is set to (from left to right) $10, 10, 40, 20$.} 
  \label{fig:SERvsSNR_MIMO_detection}
\end{figure*}%
\subsubsection{\textbf{Construction of $y_j$}}
The $j$th channel observation~${y_j \in \mathbb{R}}$ can be trivially represented as a rank-$1$ \ac{TT}. We obtain an exact \ac{TT} representation by setting ${G_1 = y_j \bm{1}_L}$ and all core tensors ${G_j = \mathbf{1}_{L}}$ for $j=2,\dots,\NR$. Note that this is equivalent to the representation 
\[
y_j  \bigtimes_{j=1}^N \mathbf{1}_{L}
 \in \mathbb{R}^{L \times \dots \times L}. 
\]

\subsubsection{\textbf{Construction of $\mathbf{h}_j^\top \bm{x}$}}
Let ${\bm{v} = (a_1,\ldots,a_L)^\top}$ be the alphabet vector with ${\mathcal{A} = \{a_1,\ldots,a_L\}}$. Further, let ${\bm{h}_j \in \R^{\NT}}$ and define ${\bm{U} = (\bm{v} , \bm{1}_L) \in \R^{L \times 2}}$. Using ${\bm{h}^\top_j = (h_j^1,\dots,h_j^{\NT})}$, $\mathbf{h}_j^\top \bm{x}$ can be represented in the \ac{TT} format of rank~$3$ with the core tensors
\begin{align*}
    G_1 &= \begin{pmatrix}
        0 & h_j^{\NT} & 0 \\ 1 & 0 & 0
    \end{pmatrix} \times_2 \bm{U} \in \R^{1 \times L \times 3}, \\
    G_{\NT} &= \begin{pmatrix}
        h_j^1 & 0 \\0 & 1 \\ 0 & 0
    \end{pmatrix} \times_2 \bm{U} \in \R^{3 \times L \times 1}.
\end{align*}
For $i=2,\dots,\NT-1$ we first define the slices
\begin{align*}
    \widetilde G_i(:,1,:) = \begin{pmatrix}
        0 & h_j^{\NT-i+1} & 0 \\ 0 & 0 & h_j^{\NT-i+1}\\ 0 & 0 & h_j^{\NT-i+1}
    \end{pmatrix}, \quad
    \widetilde G_i(:,2,:) = \bm{I}_3,
\end{align*}
and then set ${G_i = \widetilde G_i \times_2 \bm{U} \in \R^{3 \times L \times 3}}$. Although the reversed ordering of $h_j^{\NT},\ldots,h_j^1$ within $G_1,\ldots,G_{\NT}$ may appear unconventional, this construction is, to the best of our knowledge, the simplest. We note, however, that alternative constructions exist that represent the same tensor.

The addition ${y_j - \mathbf{h}_j^\top \bm{x}}$ is of rank at most~$4$, recall Sec.~\ref{subsec:TT_computations}. By performing the element-wise and scalar multiplication from Sec.~\ref{subsec:TT_computations}, we can exactly represent each of the $\NR$ log-likelihood terms~${\ell(y_j|\bm{x})=-(y_j - \mathbf{h}_j^\top \bm{x})^2/(2\sigma^2)}$ in \ac{TT} format with rank~${16}$ or lower.
For a possible rank reduction, we perform a truncation with given tolerance~$\vartheta>0$ after constructing each log-likelihood term.

Algorithm~\ref{alg:tt_det} summarizes the proposed \ac{TT}-based algorithm for symbol-wise \ac{MIMO} detection. Note that all constructions and operations in Algorithm~\ref{alg:tt_det} are performed in the \ac{TT} format, except for the hard decision in line~4, which operates on the explicit vector representation of the symbol-wise posterior marginals.
We refer to this algorithm as TTDet, and distinguish its two variants by the cross approximation method applied for the exponentiation in \ac{TT} format in line~3: TTDet-sample, based on the more classical \ac{TT}-cross algorithm, and TTDet-sweep, based on the \ac{DMRG}-like cross algorithm.
\begin{algorithm}[t]
    \DontPrintSemicolon
    \KwInit{\texttt{r\_max}, $\vartheta$}
    \KwData{Observation~$\bm{y} \in \mathbb{R}^{\NR}$, \ac{CSI}~$\left\{ \bm{H} \in \mathbb{R}^{\NR \times \NT}, \sigma^2 \right\}$}

    \algphase{TT construction}
    
    Construct $\ell(y_j|\bm{x}), \quad j=1,\ldots,\NR$ \tcp*{Sec.~\ref{subsec:mimo_log_likelihood_construction}}
    
    Compute $\Lambda(\bm{x}) = \sum_{j=1}^{\NR} \ell(y_j|\bm{x})$ \tcp*{Sec.~\ref{subsubsec:sum_over_log_likelihoods}}

    \algphase{TT exponentiation \& marginalization}
    
    Compute symbol-wise \acp{APP} \tcp*{Sec.~\ref{subsubsec:ttexp}-B4} \nonl  %
    \vspace{-1em}
    \[
    P(x_i=a|\bm{y}) \propto  \sum\limits_{\mathclap{\substack{\bm{a} \in \mathcal{A}^{\NT}\\ a_i = a}}} \mathrm{e}^{\Lambda(\bm{a})}, \quad a\in \mathcal{A}, \, i = 1,\ldots, \NT,
    \] \nl

    \algphase{Switch from TT to explicit vector format}
    
    \ac{MAP} hard decision ${\hat{x}_{i} = \arg\max\limits_{a \in \mathcal{A}} P(x_i=a | \bm{y})}$ \\
    
    \KwResult{Detection result~$\hat{\bm{x}} \gets (\hat{x}_{1}, \ldots, \hat{x}_{\NT})^\top$}
    \caption{TTDet for MIMO detection} 
    \label{alg:tt_det}
\end{algorithm}

\subsection{Numerical Evaluation} \label{subsec:mimo_results}
We numerically evaluate the performance of the TTDet algorithm for different \ac{MIMO} settings\footnote{All numerical results in this paper were obtained using the MATLAB TT-Toolbox \cite{oseledets2016_toolbox} and the MATLAB Tensor Toolbox \cite{Tensor_Toolbo_Kolda}. We provide the source code for all simulations in~\cite{schmid2026github}.}.
We consider $4$- and $16$-\ac{QAM} constellations and quadratic \ac{MIMO} channel matrices with~${\NTc=\NRc=:\tilde{N}=8,16,32}$. For these dimensions, a full tensor representation of the \ac{APP} distribution~${P(\bm{x}|\bm{y}})$ is infeasible, as it would require storage of~${M^{\tilde{N}}}$ elements.
As a baseline, we consider three well-established \ac{MIMO} detection algorithms:
\begin{itemize}
    \item The \ac{LMMSE} detector, and the \ac{EP} detector~\cite{cespedes_expectation_2014} with $10$ iterations. Both algorithms rely on a Gaussian approximation of the \ac{APP} distribution, followed by a symbol-wise nearest-neighbor decision.
    \item The sphere decoder~\cite{yang_fifty_2015}, a tree-search based MIMO detection algorithm. We employ the sphere detector without early termination, such that it achieves optimal performance in the \ac{MLSE} sense. 
\end{itemize}

Fig.~\ref{fig:SERvsSNR_MIMO_detection} shows the \ac{SER} over the \ac{SNR} for the considered \ac{MIMO} detectors.
To estimate the \ac{SER}, we perform Monte Carlo simulations in which the transmit vector~$\bm{x}$ and the channel matrix~$\bm{H}$ are randomly sampled for each new transmission, as described in Sec.~\ref{subsec:mimo_channel_model}. 
All detection algorithms are then evaluated on the same data samples. 
For each \ac{SNR} value, the simulation is continued until the sphere decoder records~$100$ block errors, i.e., transmissions in which one or more symbol errors occur.

For the $4$-\ac{QAM} scenarios, the proposed TTDet algorithm achieves near-optimal performance, closely approaching the \ac{MLSE}-optimal performance of the sphere decoder. It significantly outperforms the \ac{LMMSE} detector, and yields a $0.7$~dB gain over the \ac{EP} detector at a target ${\text{SER} = 10^{-2}}$.
The TTDet-sample variant behaves more robustly in the low-\ac{SNR} regime, where it slightly outperforms the TTDet-sweep variant and the sphere decoder in terms of \ac{SER}\footnote{Note that the sphere detector is optimal in the \ac{MLSE} sense, but not necessarily in terms of \ac{SER}, which explains the slightly lower \ac{SER} of the TTDet-sample and \ac{EP} detector for low \ac{SNR} values.}.

We emphasize that \acp{TT} with a maximum rank of~${\texttt{r\_max}=10}$ are sufficient to initialize the \ac{TT}-cross algorithm and achieve this quasi-optimal performance, indicating the low-rank nature of this problem.
Note that the \ac{DMRG}-like \ac{TT}-cross algorithm may require substantially higher intermediate ranks than~${\texttt{r\_max}=10}$.
Table~\ref{tab:memory_analysis_mimo} reports the maximum \ac{TT} rank~$r_\text{max}$ after exponentiation at the output of the \ac{TT}-cross algorithm (TTDet-sample variant) for the $4$-\ac{QAM} scenario with~${\tilde{N}=32}$.
Additionally, we analyze the memory savings relative to the explicit tensor representation, where the latter requires $M^{\tilde{N}}$ floating-point numbers to store.
In our simulations, this corresponds to a memory reduction in the order of $10^{13}$ and $10^{14}$ at $\text{SNR}=-11\,$dB and $-5\,$dB, respectively. 
\begin{table}[tb]
\centering
\caption{Memory complexity of the TTDet-sample algorithm \\for ${\tilde{N}=32}$ and $4$-\ac{QAM}}
\begin{tabular}{ r rr rr }
\toprule
 & \multicolumn{2}{c}{$r_\text{max}$ after \ac{TT}-cross} & \multicolumn{2}{c}{Memory savings w.r.t.\ $M^{\tilde{N}}$} \\
$\ebno$ & \hspace{10pt} mean & median & \hspace{30pt}mean & median \\
\midrule
$-11\,$dB & $210$ & $209$ & $10^{13}$ & $10^{13}$ \\
$-5\,$dB  & $98$  & $86$  & $10^{14}$ & $10^{14}$ \\
\bottomrule
\end{tabular}
\label{tab:memory_analysis_mimo}
\end{table}

Finally, we consider the \ac{SER} performance for the $16$-\ac{QAM} scenarios in Fig.~\ref{fig:SERvsSNR_MIMO_detection}. 
The TTDet algorithm significantly outperforms the \ac{LMMSE} detector and achieves a $2$~dB gain over the \ac{EP} detector. 
In the high-\ac{SNR} regime, the TTDec algorithm exhibits a performance degradation relative to the sphere decoder. 
Here, the TTDet-sweep variant slightly outperforms the TTDet-sample variant.
This suggests that in this regime, the \ac{TT}-cross algorithm may fail to approximate the exponential sufficiently accurate, which indicates that the Taylor-series initialization with ${\texttt{r\_max} = 20}$ and $40$, respectively, is insufficient to achieve optimal performance or higher intermediate ranks in the \ac{TT}-cross are required. 

\begin{figure}[t]
\centering
  \begin{tikzpicture}
  \begin{axis}[
    width=0.6\columnwidth, %
    height=0.6\linewidth,
    align = left,
    grid=major, %
    grid style={gray!30}, %
    xlabel= SNR (dB),
    xlabel style={yshift=3pt, xshift=0pt},
    ylabel= SER,
    ymode=log,
    ymax = 0.1,
    ymin = 0.001,
    xmin = 6,
    xmax = 10,
    enlarge x limits=false,
    enlarge y limits=false,
    line width=1pt,
	  legend style={font=\footnotesize, cells={align=left}, draw=gray, anchor=south west, at={(0.04,0.03)}},
    legend cell align={left},
    mark options={solid, mark size=1.1pt},
    every axis plot/.append style={line width=1.1pt },
    colormap name=coolsafe,
    cycle list={[of colormap=coolsafe, samples=8]},
    colorbar,
    colorbar style={
        title={$\texttt{r\_max}$},
        title style={at={(3,0.45)}, anchor=west, xshift=0pt},
        ytick={0, 200, 400, 600, 800, 1000},
        yticklabels={6, 8, 12, 16, 20, 30},
        y dir=reverse,
        height=0.42\linewidth,
        width=0.75em,
        at={(1.1, 0.0)},
        anchor=south west,
    },
point meta min=0,
point meta max=1000,
    legend image code/.code={
  \draw[#1] (0cm,0cm) -- (0.27cm,0cm);
},
    ]
\addplot[draw=none, light_gray_legend] table[x={SNR (dB)}, y={SER}, col sep=comma]
    {numerical_results/M4_N16_rmax6_results_tt_multifuncrs.csv};
\addplot[draw=none, dotted, light_gray_legend] table[x={SNR (dB)}, y={SER}, col sep=comma] {numerical_results/M4_N16_results_sphere_decoder.csv};
\legend{TTDet-sample \\ Sphere decoder \\}

\addplot[ color of colormap={1000 of coolsafe}] table[x={SNR (dB)}, y={SER}, col sep=comma]
    {numerical_results/M4_N16_rmax30_results_tt_multifuncrs.csv};
\addplot[ color of colormap={800 of coolsafe}] table[x={SNR (dB)}, y={SER}, col sep=comma]
    {numerical_results/M4_N16_rmax20_results_tt_multifuncrs.csv};
\addplot[ color of colormap={600 of coolsafe}] table[x={SNR (dB)}, y={SER}, col sep=comma]
    {numerical_results/M4_N16_rmax16_results_tt_multifuncrs.csv};
\addplot[ color of colormap={400 of coolsafe}] table[x={SNR (dB)}, y={SER}, col sep=comma]
    {numerical_results/M4_N16_rmax12_results_tt_multifuncrs.csv};
 \addplot[ color of colormap={200 of coolsafe}] table[x={SNR (dB)}, y={SER}, col sep=comma]
    {numerical_results/M4_N16_rmax8_results_tt_multifuncrs.csv};   
\addplot[ color of colormap={0 of coolsafe}] table[x={SNR (dB)}, y={SER}, col sep=comma]
    {numerical_results/M4_N16_rmax6_results_tt_multifuncrs.csv};
    
\addplot[M4N16_color_overlay, dotted] table[x={SNR (dB)}, y={SER}, col sep=comma] {numerical_results/M4_N16_results_sphere_decoder.csv};
\end{axis}
\end{tikzpicture}
\caption{\ac{SER} over \ac{SNR} for different ranks~\texttt{r\_max} of the truncated Taylor series initialization of the TTDet-sample algorithm for a \ac{MIMO} system with ${\tilde{N}=8}$ and $16$-\ac{QAM}.}
\label{fig:rank_sweep_mimo}
\end{figure}
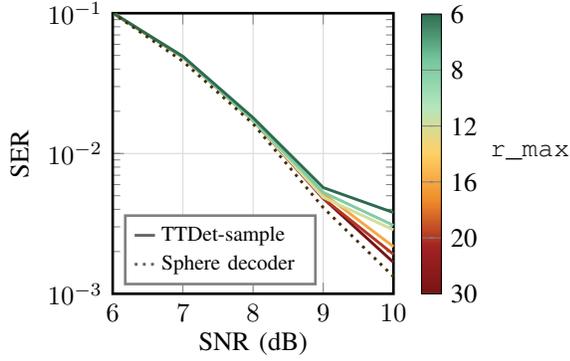

Fig.~\ref{fig:rank_sweep_mimo} illustrates the dependency of the \ac{SER} performance on the maximum rank~\texttt{r\_max} of the truncated Taylor series initialization for ${\tilde{N}=8}$.
At low \ac{SNR}, a small rank~${\texttt{r\_max}=6}$ suffices to achieve quasi-optimal performance. 
In the high-SNR regime, we can observe a clear convergence behavior as \texttt{r\_max} increases. This indicates that higher ranks are required to close the remaining gap to optimality.

\section{Example: Decoding of Linear Channel Codes}\label{sec:cc}
Error-correction coding is a fundamental component of modern digital communication systems, enabling reliable data transmission over noisy channels~\cite{Sha48} and forming the basis of many wireless, storage, and optical communication standards. 
Efficient channel decoding algorithms are essential to approach theoretical performance limits while remaining computationally feasible in practical implementations.
In this section, we consider the problem of decoding linear block codes~\cite{richardson_modern_2008}. 
\subsection{Linear Coding over the BI-\ac{AWGN} Channel}
Consider the problem of reliably transmitting an information word~${\bm{u} \in \mathbb{F}_2^k}$ over a noisy communication channel, where each bit $u_i$ is independently and uniformly sampled from the finite field ${\mathbb{F}_2 := \{0,1\}}$.
To introduce redundancy and enable error correction capabilities, 
a binary linear block encoder maps each of the $2^k$ information words~${\bm{u}}$ bijectively to a codeword
\[
    \bm{c} = \bm{G} \bm{u}, \qquad \bm{G}\in \mathbb{F}_2^{n\times k},
\]
using the generator matrix~$\bm{G}$. 
The set of all~$2^k$ codewords forms a linear subspace of~$\mathbb{F}_2^n$ and defines the binary linear block code ${\mathcal{C}(n,k)}$ with rate~${R:=k/n<1}$.

We consider the \ac{BI-AWGN} channel, where each codeword~$\bm{c}$ is mapped to a sequence of \ac{BPSK} symbols~$\bm{x}$ with ${x_j = (-1)^{c_j}}, {j=1,\ldots,n}$ and transmitted over an \ac{AWGN} channel
\[
    \bm{y} = \bm{x} + \bm{n}, \qquad \bm{n} \sim \mathcal{N}\mleft(\bm{0}, \frac{N_0}{2} \bm{I}_{n}\mright),
\]
where we define the \ac{SNR} in terms of the energy per information bit
\[
E_\text{b}/N_0 := -10 \log_{10} \mleft( r N_0 \mright) \quad \text{in dB}.
\]

Channel decoding aims to reconstruct the information word~$\bm{u}$ based on the observation~${\bm{y} \in \mathbb{R}^n}$ at the receiver.
It can be formulated as a detection problem in the additive noise model defined in Sec.~\ref{subsec:additive_noise_channel_model} with~${\NR=n}$, ${\NT=k}$. The corresponding \ac{APP} is given by
\begin{align*}
    P(\bm{u}|\bm{y}) 
    &\propto p(\bm{y}|\bm{u}) P(\bm{u}) = \sum\limits_{\bm{c} \in \mathbb{F}_2^n} p(\bm{y}|\bm{u}, \bm{c}) P(\bm{\bm{c}|\bm{u}}) P(\bm{u}) \\
    &\propto  \sum\limits_{\bm{c} \in \mathbb{F}_2^n} \prod\limits_{j=1}^{n} \exp \mleft( -\frac{(y_j-(-1)^{c_j})^2}{N_0} \mright) \mathbbm{1}_{\left\{\bm{c} = \bm{G} \bm{u}\right\}},
\end{align*}
or, equivalently, in the logarithmic domain
\begin{align}
    \log P(\bm{u}|\bm{y}) &= C - \frac{1}{N_0}\sum\limits_{j=1}^n \left( y_j - (-1)^{( \bm{G} \bm{u})_j} \right)^2, \nonumber \\
    &= C + \sum\limits_{j=1}^n \ell(y_j | \bm{u}) =: C + \Lambda(\bm{u}),
    \label{eq:cc_app}
\end{align}
where~$C$ is a constant independent of $\bm{u}$.

\subsection{TT Construction of the Log-APP} \label{subsec:cc_log_APP_construction}
Our objective is to express the unnormalized log-\ac{APP} metric~${\Lambda(\bm{u})}$ in~\eqref{eq:cc_app} in the \ac{TT} format.
Although the code is linear, the decoding problem does not correspond to the linear observation model in~\eqref{eq:MIMO_model}. This is because the linearity of the code ${\bm{c} = \bm{G} \bm{u}}$ is defined over~$\mathbb{F}_2$, rather than over~$\R$.
Therefore, we incorporate the code constraint directly into the \ac{TT} construction. Specifically, we construct a single \ac{TT} representation of~${\Lambda(\bm{u})}$ rather than forming the~${\NR=n}$ log-likelihood terms~$\ell(y_j|\bm{u})$ individually and summing their corresponding \acp{TT} representations afterwards.

We expand the quadratic term of~\eqref{eq:cc_app} into
\begin{equation} \label{eq:log_app_cc}
    \Lambda(\bm{u}) = - \frac{1}{N_0} \sum\limits_{j=1}^n \! \left( y_j^2 + 1 - 2 y_j \, \prod\limits_{\substack{i=1\\G_{ji}=1}}^k \, (-1)^{u_i} \right).
\end{equation}
The term $- \frac{1}{N_0}\sum_{j=1}^{n} ( y_j^2 +1)$ in~\eqref{eq:log_app_cc} is constant with respect to~$\bm{u}$ and can be omitted in the context of decoding, as discussed in Sec.~\ref{subsec:additive_noise_channel_model}. Each summand 
\[
    \frac{2y_j}{N_0} \prod\limits_{\substack{i=1\\G_{ji}=1}}^k (-1)^{u_i}, \quad j=1,\dots,n
\]
is a scalar function that admits a rank-$1$ \ac{TT} representation. 
Consequently, the sum over the $n$ scalar terms can be expressed by a \ac{TT} of rank at most~$n$. 
For the construction, we define 
\begin{align*}
    s_j^i := 
\begin{cases}
-1 & \text{if } G_{ji}=1,\\
+1  & \text{otherwise}.
\end{cases}
\end{align*}
The first core $G_1$ and the last core $G_{\NT}$ have the form
\begin{align}
    G_1 &= \frac{2}{N_0} \begin{pmatrix}
    y_1s_1^1 & y_2s_2^1 & \cdots& y_{n} s_{n}^1 \\
    y_1 & y_2 &\cdots& y_{n}
    \end{pmatrix} \in \R^{1 \times 2 \times n} , \label{eq:G1_cc} \\
    G_{k} &= \begin{pmatrix}
        s_1^{k} & s_2^{k} & \cdots& s_{n}^{k} \nonumber \\
        1&1&\cdots& 1
    \end{pmatrix}^\top  \in \R^{{n} \times 2 \times 1}.
\end{align}
The remaining cores ${G_i \in \R^{n\times 2 \times n}}$, ${i=2,\dots,k-1}$, can be constructed by
\begin{align*}
    G_i(:,1,:) &= \text{diag}\left( (s_1^i, s_2^i,\dots,s_{n}^i)^\top  \right), \\
    G_i(:,2,:) &= \text{diag}\left( \bm{1}_n  \right).
\end{align*}
Note that the construction of the cores ${G_2,\ldots,G_{k}}$ is independent of the observation~$\bm {y}$. Hence, for a given code, only $G_1$ needs to be updated when a new~$\bm{y}$ is received.

The presented construction is the simplest one we found, although it is not optimal in terms of rank. 
However, applying a standard truncation with a small tolerance (e.g., $\vartheta = 10^{-12}$) yields a \ac{TT} representation with smaller ranks.

Based on the \ac{TT} construction of the log-\ac{APP} term in Sec.~\ref{subsec:cc_log_APP_construction}, followed by exponentiation and marginalization as described in Sec.~\ref{subsec:main_contribution}, we obtain symbol-wise \acp{APP}.
The decoded information word~${\hat{\bm{u}}}$ is then obtained via the \ac{MAP} hard decision in~\eqref{eq:max}. 

\subsection{Adaptive Rank Decoding} \label{subsec:adaptive_rank_decoding}
The rank required to reliably represent the joint \ac{APP} distribution after exponentiation varies significantly across transmissions, as it strongly depends on the specific noise realization. 
The \ac{TT}-cross algorithm inherently finds low-rank representations and adapts intermediate ranks during the approximation procedure, see Sec.~\ref{subsubsec:ttexp}.
Nevertheless, certain hyperparameters must be selected in advance, in particular the maximum rank~\texttt{r\_max} of the truncated Taylor-series initialization.
Configuring this parameter for the worst case scenario introduces unnecessary computational overhead in the majority of transmissions, where lower ranks are sufficient.

We therefore propose an adaptive decoding strategy that operates with reduced complexity in the low-rank regime and increases~\texttt{r\_max} only when required. 
An additional consideration is the stochastic nature of the \ac{TT}-cross algorithm. Due to the random initialization of the inflated ranks in each sweep, repeated runs may yield different results. Both aspects motivate a statistical test that assesses whether the current decoding result is sufficiently reliable. 

In general, verifying whether a candidate corresponds to the \ac{ML} solution is as hard as solving the original detection problem.
Channel decoding, however, offers a structural advantage: the observation~$\bm{y}$ is a noisy version of a codeword.
Classical decoders exploit this by checking whether a candidate~${\hat{\bm{x}} \in \mathbb{F}_2^n}$ satisfies the code constraints, i.e., ${\hat{\bm{x}} \in \mathcal{C}(n,k)}$~\cite{richardson_modern_2008}.
In the proposed approach, we instead estimate the information word~$\bm{u}$ directly, and every vector in~$\mathbb{F}^k$ is a valid candidate, i.e., code membership cannot serve as a reliability indicator.
Instead, we leverage the known noise statistics together with structural properties of the code, in particular its minimum Hamming distance~$d_{\min}$.

Specifically, we consider the squared Euclidean distance 
\[
    d(\hat{\bm{x}},\bm{y}) = \sum_{j=1}^n \left( y_j - \hat{x}_j \right)^2, \quad \hat{x}_j = (-1)^{\hat{c}_j}, \; \hat{\bm{c}} = \bm{G}\hat{\bm{u}}
\]
between the re-encoded and \ac{BPSK}-modulated candidate~${\hat{\bm{x}}}$ and the observation~$\bm{y}$, and formulate a binary hypothesis test:
\begin{itemize}
    \item ${H_0\colon}$ \textit{correct decoding}, $\hat{\bm{c}} = \bm{c} \Leftrightarrow d_\text{H}(\hat{\bm{c}},\bm{c})=0$,
\end{itemize}
    \begin{itemize}
    \item ${H_1}\colon$ \textit{decoding to incorrect codeword}, $d_\text{H}(\hat{\bm{c}},\bm{c}) = d_{\min}$.
\end{itemize}
For simplicity, $H_1$ only considers decoding to an incorrect codeword at the minimum Hamming distance $d_{\min}$, which represents the dominant decoding error event at high~\ac{SNR}.
Under~$H_0$, 
the squared Euclidean distance reduces to the squared noise norm~${d(\hat{\bm{x}},\bm{y}) = \|\bm{n}\|^2}$, which follows a scaled chi-squared distribution $\frac{N_0}{2}\chi^2_n$ with $n$ degrees of freedom. 
Under~$H_1$, the $d_{\min}$ positions where $\hat{\bm{c}}$ and $\bm{c}$ differ each contribute a shifted noise term. 
Consequently, ${d(\hat{\bm{x}},\bm{y})}$ follows a scaled noncentral chi-squared distribution~$\frac{N_0}{2}\chi^2_n(\lambda)$ with noncentrality parameter~$\lambda = \frac{8 d_{\min}}{N_0}$.
Let $f_{H_0}$ and $f_{H_1}$ denote the probability density functions of ${d(\hat{\bm{x}},\bm{y})}$ under $H_0$ and $H_1$, respectively.

Since both hypotheses are simple, i.e., $f_{H_0}$ and $f_{H_1}$ are fully specified with known $n$, $N_0$, and $d_{\min}$, the Neyman-Pearson lemma guarantees that the likelihood-ratio test
\begin{equation*}
    T\mleft(d(\hat{\bm{x}},\bm{y})\mright) = \frac{f_{H_1}\mleft(d(\hat{\bm{x}},\bm{y})\mright)}{f_{H_0}\mleft(d(\hat{\bm{x}},\bm{y})\mright)} 
    \overset{H_1}{\underset{H_0}{\gtrless}} \eta
\end{equation*}
is uniformly most powerful~\cite{casella2002statistical}. 
Since the likelihood ratio is monotonic in~${d(\hat{\bm{x}},\bm{y})}$, the test reduces to a simple threshold test on this quantity.
If the test selects~$H_1$, we re-execute the decoder with incremented~\texttt{r\_max}; otherwise the result is accepted and the algorithm stops early.

\begin{algorithm}[t]
    \DontPrintSemicolon
    \KwInit{$\bm{G}, \vartheta, \nu^\star \! \gets \! \infty$,
    $( \texttt{\footnotesize r\_max}^{(1)},\ldots,{ \texttt{\footnotesize r\_max}^{(t)}})$}
    \algphase{Offline preprocessing}
    Construct \ac{TT} cores~$G_2,\ldots,G_{k}$ of~$\Lambda(\bm{u})$ \hspace{-1em} \tcp*{Sec.~\ref{subsec:cc_log_APP_construction}}
    \algphase{Online per-observation processing}
    \KwData{Observation~$\bm{y} \in \mathbb{R}^{n}$, \ac{CSI}~$\left\{ N_0 \right\}$}

    Construct $G_1$ to obtain full \ac{TT} of~$\Lambda(\bm{u})$\tcp*{Eq.~\eqref{eq:G1_cc}}

    Evaluate $\tilde{P}_\text{e}$ and $\eta ( \tilde{P}_\text{e} )$ \tcp*{Eqs.~\eqref{eq:eta_calculation} and \eqref{eq:normal_approx}}

    \For(\tcp*[f]{Sec.~\ref{subsec:adaptive_rank_decoding}}){$\texttt{\footnotesize\upshape r\_max} = \texttt{\footnotesize\upshape r\_max}^{(1)},\ldots,\texttt{\footnotesize\upshape r\_max}^{(t)}$}{
    \algphase{TT exponentiation \& marginalization}
    Compute symbol-wise \acp{APP} \tcp*{Sec.~\ref{subsubsec:ttexp}-B4} \nonl  %
    \vspace{-1em}
    \[
    P(u_i=a|\bm{y}) \propto  \sum\limits_{\mathclap{\substack{\bm{a} \in \mathbb{F}_2^k,  a_i = a}}} \mathrm{e}^{\Lambda(\bm{a})}, \quad a\in \mathbb{F}_2, \, i = 1,\ldots, k,
    \] \nl
    
    \algphase{Switch from TT to explicit vector format}
    \ac{MAP} hard decision ${\hat{u}_{i} = \arg\max\limits_{a \in \mathbb{F}_2^k} P(u_i=a | \bm{y})}$ \\

    Evaluate $\nu_\text{cand} = d\mleft((-1)^{\bm{G}\hat{\bm{u}}},\bm{y}\mright), \,\hat{\bm{u}} = (\hat{u}_{1},\ldots,\hat{u}_{k})^\top$ \hspace{-1em}

    \If{$\nu_\text{cand} < \nu^\star$}{
        $\hat{\bm{u}}^\star \gets \hat{\bm{u}}, \quad \nu^\star \gets \nu_\text{cand}$ \\
        \lIf{$\nu^\star \! < \! \eta ( \tilde{P}_\text{e} )\!$}{break \tcp{\hspace{-2pt}$\hspace{-0.5pt} H_0 \colon$ \hspace{-10pt} early \hspace{-6pt} stopping}} \vspace{-1em}
    }
    }
    \KwResult{Decoding result~$\hat{\bm{u}} \gets \hat{\bm{u}}^\star$}
    \caption{TTDec for decoding of linear codes} 
    \label{alg:tt_dec}
\end{algorithm}
We choose the threshold~$\eta$ to set the type-II error probability, i.e., the probability that the decoder terminates early although the candidate is incorrect,
\begin{align*}
    P\left(\text{accept } H_0 \, | \, H_1 \right) 
    &= P\left( d(\hat{\bm{x}},\bm{y}) < \eta \, | \, H_1 \right) \\
    &= F_{\chi^2_n(\lambda)}\mleft( \frac{2 \eta}{N_0} \mright)    
    \stackrel{!}{=} \frac{\tilde{P}_\text{e}}{100},
\end{align*}
relative to the minimum achievable block error probability~$\tilde{P}_\text{e}$.
Here, $F_{\chi^2_n(\lambda)}$ denotes the cumulative distribution function of the noncentral chi-square distribution.
Solving for $\eta$ leads to the threshold
\begin{align}
    \eta ( \tilde{P}_\text{e} ) = \frac{N_0}{2} F_{\chi^2_n(\lambda)}^{-1} \mleft( \frac{\tilde{P}_\text{e}}{100} \mright). \label{eq:eta_calculation}
\end{align}
To approximate~$\tilde{P}_\text{e}$ for a given code of rate~$R$ and blocklength~$n$, we use the refined normal approximation from finite-blocklength information theory~\cite{polyanskiy2010channel, durisi20-11a}:
\begin{equation}
    \tilde{P}_\text{e} := Q\!\left( \frac{C - R + \frac{\log n}{2n}}
    {\sqrt{V/n}} \right), \label{eq:normal_approx}
\end{equation}
where $C$ denotes the channel capacity and $V$ the channel dispersion, as defined in~\cite{durisi20-11a}.

Algorithm~\ref{alg:tt_dec} summarizes the proposed \ac{TT}-based algorithm for bit-wise channel decoding of linear codes over the \ac{BI-AWGN} channel.
The offline phase precomputes the \ac{TT} cores ${G_2,\ldots,G_k}$ once per code, while the online phase processes each observation~$\bm{y}$ by constructing~$G_1$, evaluating the early stopping threshold~$\eta ( \tilde{P}_\text{e} )$, and iteratively increasing the rank~\texttt{r\_max} of the truncated Taylor-series intialization until either the stopping criterion is met or the maximum rank is reached.
Note that the constructions and operations in lines~1, 2, and 5 are performed entirely in the \ac{TT} format.
We refer to this algorithm as TTDec, and distinguish its two variants by the cross approximation method used for the exponentiation: TTDec-sample, based on the more classical \ac{TT}-cross algorithm, and TTDec-sweep, based on the \ac{DMRG}-like cross algorithm.

\subsection{Numerical Evaluation}

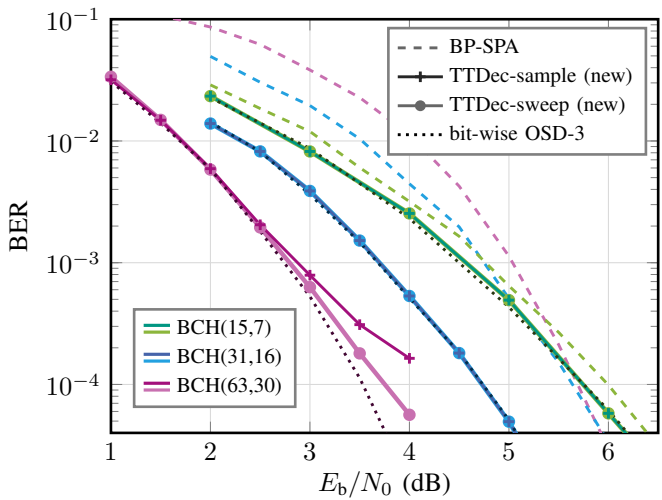
\begin{figure}[t]
\centering
  \begin{tikzpicture}
  \begin{axis}[
    width=\columnwidth, %
    height=0.8\linewidth,
    align = left,
    grid=major, %
    grid style={gray!30}, %
    xlabel= $\ebno$ (dB),
    xlabel style={yshift=3pt, xshift=0pt},
    ylabel= BER,
    ymode=log,
    ymax = 0.1,
    ymin = 0.00004,
    xmin = 1,
    xmax = 6.5,
    enlarge x limits=false,
    enlarge y limits=false,
    line width=1pt,
	  legend style={font=\footnotesize, cells={align=left}, draw=gray, anchor=south west, at={(0.04,0.06)}},
    legend cell align={left},
    mark options={solid, mark size=1.1pt},
    every axis plot/.append style={line width=1.1pt },
    axis line style={line width=1pt},
    axis background/.style={fill=white},
    ]
\addlegendimage{
legend image code/.code={
\draw[15_7_color,line width=1.2pt] (0cm,-0.04cm) -- (0.35cm,-0.04cm);
\draw[15_7_color_multifuncrs,line width=1.2pt] (0cm,0.05cm) -- (0.35cm,0.05cm);
}}
\addlegendentry{BCH(15,7)}
\addlegendimage{
legend image code/.code={
\draw[31_16_color,line width=1.2pt] (0cm,-0.04cm) -- (0.35cm,-0.04cm);
\draw[31_16_color_multifuncrs,line width=1.2pt] (0cm,0.05cm) -- (0.35cm,0.05cm);
}}
\addlegendentry{BCH(31,16)}
\addlegendimage{
legend image code/.code={
\draw[63_30_color,line width=1.2pt] (0cm,-0.04cm) -- (0.35cm,-0.04cm);
\draw[63_30_color_multifuncrs,line width=1.2pt] (0cm,0.05cm) -- (0.35cm,0.05cm);
}}
\addlegendentry{BCH(63,30)}

\addplot[draw=63_30_color, dashed] coordinates {
 (1.00,  1.284471e-01)
 (1.50,  1.071429e-01)
 (2.00,  8.629248e-02)
 (2.50,  6.207914e-02)
 (3.00,  3.829688e-02)
 (3.50,  2.272618e-02)
 (4.00,  1.122838e-02)
 (4.50,  4.249815e-03)
 (5.00,  1.131561e-03)
 (5.50,  2.121770e-04)
 (6.00,  2.960132e-05)
};
\addplot[63_30_color, mark=*, mark options={mark size=1.5pt}, line width=1.7pt] coordinates {(1,0.0336666) (1.5,1.49e-02) (2,0.0058222) (2.5,1.95e-03)(3,6.3e-04) (3.5,1.8e-04) (4,5.6344086e-05) };
\addplot[63_30_color_multifuncrs, mark=+, mark options={mark size=2pt}] coordinates {(1,0.03181) (1.5,1.48e-02) (2,0.0059222) (2.5,2.05e-03)(3,7.9e-04) (3.5,3.1e-04) (4,1.639086e-04) };
\addplot[draw=63_30_color_overlay, dotted] coordinates {(1,3.08e-02) (1.5,1.40e-02) (2,5.88e-03) (2.5,1.81e-03) (3,5.28e-04) (3.5,1.13e-04) (4,1.55e-05)};

\addplot[draw=31_16_color, dashed] coordinates {
 (2.00,  4.952077e-02)
 (2.50,  3.035680e-02)
 (3.00,  1.953834e-02)
 (3.50,  1.041309e-02)
 (4.00,  4.448110e-03)
 (4.50,  1.968986e-03)
 (5.00,  5.083567e-04)
 (5.50,  1.518205e-04)
 (6.00,  3.378002e-05)
};
\addplot[31_16_color, mark=*, mark options={mark size=1.5pt}, line width=1.7pt] coordinates {(2,0.0139)(2.5,8.21e-03)(3,0.003887)(3.5,1.52e-03)(4,0.000533)(4.5,1.81e-04)(5,4.948e-05)(5.5,9.85e-06)};
\addplot[31_16_color_multifuncrs, mark=+, mark options={mark size=2pt}]  coordinates {(2,0.0139)(2.5,8.21e-03)(3,0.003887)(3.5,1.52e-03)(4,0.000533)(4.5,1.81e-04)(5,4.948e-05)(5.5,9.85e-06)};
\addplot[draw=31_16_color_overlay, dotted] coordinates {
(2,1.43e-02)
(2.5,8.15e-03)
(3,3.63e-03)
(3.5,1.47e-03)
(4,5.11e-04)
(4.5,1.83e-04)
(5,5.11e-05)
(5.5,9.83e-06)
};

\addplot[draw=15_7_color, dashed] coordinates {
 (2.00,  2.887961e-02)
 (2.50,  1.847099e-02)
 (3.00,  1.196314e-02)
 (3.50,  6.040709e-03)
 (4.00,  3.181271e-03)
 (4.50,  1.648341e-03)
 (5.00,  6.448736e-04)
 (5.50,  2.677752e-04)
 (6.00,  9.890811e-05)
 (6.50,  3.090116e-05)
};
\addplot[15_7_color, mark=*, mark options={mark size=1.5pt}, line width=1.7pt] coordinates {(2,0.02335)(3,0.0082)(4,0.00254)(5,0.0004927)(6,0.0000579)(6.5,1.98e-05)};
\addplot[15_7_color_multifuncrs, mark=+, mark options={mark size=2pt}] coordinates {(2,0.02335)(3,0.0082)(4,0.00254)(5,0.0004927)(6,0.0000579)(6.5,1.98e-05)};
\addplot[draw=15_7_color_overlay, dotted] coordinates {(2,2.27e-02)(3,8.67e-03)(4,2.310e-03)(5,4.31e-04)(6,6.16e-05)(6.5,1.99e-05)};

\end{axis}
 \begin{axis}[
    width=\columnwidth, %
    height=0.7\linewidth,
    align = left,
    grid style={gray!30}, %
    xlabel= $\ebno$ (dB),
    xlabel style={yshift=3pt, xshift=0pt},
    ylabel= BLER,
    ymode=log,
    ymax = 0.1,
    ymin = 0.00004,
    xmin = 1,
    xmax = 6.5,
    enlarge x limits=false,
    enlarge y limits=false,
    line width=1pt,
	  legend style={font=\footnotesize, cells={align=left}, draw=gray, anchor=north east, at={(0.98,1.17)}},
    legend cell align={left},
    mark options={solid, mark size=1.1pt},
    every axis plot/.append style={line width=1.1pt },
    axis lines=none,
    xtick=\empty,
    ytick=\empty,
    xticklabels=\empty,
    yticklabels=\empty,
axis line style={line width=1pt},
    ]
\addlegendimage{light_gray_legend, dashed}
\addlegendentry{BP-SPA}
    
\addlegendimage{medium_gray_legend, solid, mark=+, mark options={mark size=2pt}}
\addlegendentry{TTDec-sample (new)}

\addlegendimage{light_gray_legend, solid, mark=*, mark options={mark size=1.5pt}}
\addlegendentry{TTDec-sweep (new)}

\addlegendimage{dark_gray_legend, dotted}
\addlegendentry{bit-wise OSD-3}
    \end{axis}
\end{tikzpicture}
\caption{BER over $\ebno$ for various decoding algorithms across different BCH codes. The maximum rank in the Taylor-series initialization was set to ${\texttt{r\_max}=30}$ for the BCH$(63,30)$ code and ${\texttt{r\_max}=10}$ for the other codes.}
\label{fig:ber_over_ebno_cc}
\end{figure}
We evaluate the performance of the TTDec algorithm for different linear block codes. 
Fig.~\ref{fig:ber_over_ebno_cc} shows the \ac{BER} over $\ebno$ for \ac{BCH} codes of different block lengths~$n$ and code rates~${R\approx 0.5}$.
We estimate the \ac{BER} via Monte Carlo simulations. 
For each $\ebno$ value, we run simulations until at least $100$ block errors (i.e., transmissions in which one or more bit errors occur) are accumulated for both variants of the TTDec algorithm. 
To ensure a fair comparison, we use the identical noise realizations for the TTDec-sample and TTDec-sweep algorithms. 
We compare the results against \ac{OSD} with order~$3$. 
To obtain bit-wise decisions, we compute approximate marginals from the list of candidate codewords~$\mathcal{L}$ produced by \ac{OSD} and select the most likely bit value according to
\begin{equation*}
    \hat{c}_{i,\text{OSD}} = \arg\max\limits_{a \in \{0,1\}} \sum\limits_{\substack{\bm{c} \in \mathcal{L} \\ c_i=a}} p(\bm{y} | \bm{c}), \quad i=1,\ldots,n,
\end{equation*}
which serves as a near-optimal bit-wise decoding baseline for the considered codes.
As a low-complexity baseline, we employ a \ac{BP} decoder with the \ac{SPA} update rule, using up to~$1000$ parallel message-passing iterations and early stopping upon detection of a valid codeword. 
For the latter, we use the sparse ``$1$-min'' parity check matrix provided in~\cite{channelcodes}.

For the \ac{BCH}$(15,7)$ and \ac{BCH}$(31,16)$ codes, both TTDec variants achieve a \ac{BER} performance comparable to \ac{OSD}-3 across the entire evaluated $\ebno$ range.
For the \ac{BCH}$(63,30)$ code, a small performance gap to \ac{OSD}-3 becomes visible in the high-\ac{SNR} regime.
Consistent with the behavior observed for the TTDet algorithm in \ac{MIMO} detection (Sec.~\ref{subsec:mimo_results}), the TTDec-sweep algorithm achieves a lower \ac{BER} than the TTDec-sample variant. 
Compared to \ac{BP}-based decoding, we can observe a significant gain of $2\,$dB for a target ${\text{BER}=10^{-4}}$.

\begin{figure}[tb]
    \centering
  \begin{tikzpicture}
\begin{axis}[
    height=0.7\columnwidth,
    width=\columnwidth,
    ymode=log,
    log origin=infty,
    ymin=1e-3,
    ymax=1,
    xmin=0,
    xmax=1120,
    xlabel={$r_{\text{max}}$},
    ylabel={frequency},
    tick align=inside,      %
ytick align=outside,    %
axis y line*=left,
    x tick label style={/pgf/number format/1000 sep={}},
    legend style={at={(0.6,0.85)},anchor=north east, draw=gray,
    legend cell align=left,
    },
    axis line style={line width=1pt},
    legend image code/.code={
  \draw[#1] (0cm,0cm) -- (0.35cm,0cm);
},
axis background/.style={fill=white},
]
\addplot [fill=none, draw=ebno2_hist,   line width=0.7pt]
    table [x=x, y=c2, col sep=comma]
    {numerical_results/hist_data.csv} \closedcycle;
\addlegendentry{$\ebno = 2$\,dB}

\addplot [fill=none, draw=ebno3_hist,  line width=0.7pt]
    table [x=x, y=c3, col sep=comma]
    {numerical_results/hist_data.csv} \closedcycle;
\addlegendentry{$\ebno = 3$\,dB}

\addplot [fill=none, draw=ebno4_hist, line width=0.7pt]
    table [x=x, y=c4, col sep=comma]
    {numerical_results/hist_data.csv} \closedcycle;
\addlegendentry{$\ebno = 4$\,dB}

    \addplot[ebno2_hist, line width=0.7pt] coordinates {(332,0.00001) (332,10)};
    \addplot[ebno3_hist, line width=0.7pt] coordinates {(71,0.00001) (71,10)};
    \addplot[ebno4_hist, line width=0.7pt] coordinates {(17,0.00001) (17,10)};
    \addplot[ebno2_hist, line width=0.7pt, dashed, dash pattern=on 2pt off 2pt] coordinates {(140,0.00001) (140,10)};
    \addplot[ebno3_hist, line width=0.7pt, dashed, dash pattern=on 2pt off 2pt] coordinates {(14,0.00001) (14,10)};
    \addplot[ebno4_hist, line width=0.7pt, dashed, dash phase=1.5pt, dash pattern=on 2pt off 2pt] coordinates {(12,0.00001) (12,10)};

\addplot [fill=white,  draw=none]
    table [x=x, y=c2, col sep=comma]
    {numerical_results/hist_data.csv} \closedcycle;
\addplot [fill=white,  draw=none]
    table [x=x, y=c3, col sep=comma]
    {numerical_results/hist_data.csv} \closedcycle;
\addplot [fill=white,  draw=none]
    table [x=x, y=c4, col sep=comma]
    {numerical_results/hist_data.csv} \closedcycle;

\addplot [fill=ebno2_hist,   fill opacity=0.5,  draw=none]
    table [x=x, y=c2, col sep=comma]
    {numerical_results/hist_data.csv} \closedcycle;
\addplot [fill=ebno3_hist,  fill opacity=0.5,  draw=none]
    table [x=x, y=c3, col sep=comma]
    {numerical_results/hist_data.csv} \closedcycle;
\addplot [fill=ebno4_hist, fill opacity=0.5,  draw=none]
    table [x=x, y=c4, col sep=comma]
    {numerical_results/hist_data.csv} \closedcycle;

\addplot [fill=none, draw=ebno2_hist,   line width=0.7pt]
    table [x=x, y=c2, col sep=comma]
    {numerical_results/hist_data.csv} \closedcycle;

\addplot [fill=none, draw=ebno3_hist,  line width=0.7pt]
    table [x=x, y=c3, col sep=comma]
    {numerical_results/hist_data.csv} \closedcycle;

\addplot [fill=none, draw=ebno4_hist, line width=0.7pt]
    table [x=x, y=c4, col sep=comma]
    {numerical_results/hist_data.csv} \closedcycle;

\end{axis}
\begin{axis}[
    height=0.7\columnwidth,
    width=\columnwidth,
    ymode=log,
    log origin=infty,
    ymin=1e-3,
    ymax=1,
    xmin=0,
    xmax=1120,
    tick align=inside,      %
    axis y line*=right,
    x tick label style={/pgf/number format/1000 sep={}},
    legend style={at={(0.61,0.85)},anchor=north west, draw=gray,
    legend cell align=left,
    },
    xticklabels=\empty,
    yticklabels=\empty,
    yticklabel pos=right,
    axis line style={line width=1pt},
]
\addlegendimage{
legend image code/.code={
  \draw[medium_gray_legend, line width=0.7pt, fill=light_gray_legend!30] 
        (0cm,-0.05cm) -- (0.35cm,-0.05cm) -- (0.35cm,0.05cm) -- (0cm,0.05cm) -- cycle;
}}
\addlegendentry{Distribution}

\addlegendimage{
legend image code/.code={
\draw[medium_gray_legend,line width=0.7pt] (0cm,0.0cm) -- (0.35cm,0.0cm);
}}
\addlegendentry{Mean}

\addlegendimage{
legend image code/.code={
\draw[medium_gray_legend,line width=0.7pt, dashed, dash pattern=on 2pt off 2pt] (0cm,0.0cm) -- (0.35cm,0.0cm);
}}
\addlegendentry{Median}
\end{axis}

\end{tikzpicture}
    \caption{Histogram of the maximum \ac{TT} rank $r_\text{max}$ after \ac{TT}-cross approximation of the joint \ac{APP} mass function for the TTDet-sweep algorithm applied to the BCH$(63,30)$ code at different $\ebno$ values.}
    \label{fig:cc_rank_histogram}
\end{figure}
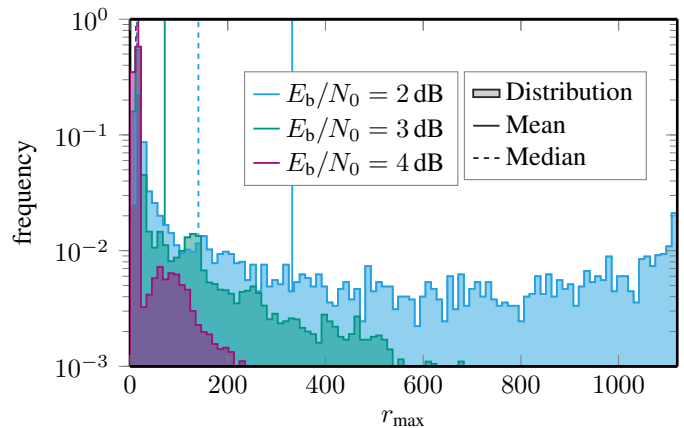

We further analyze the maximum ranks~$r_\text{max}$ encountered after the \ac{TT}-cross approximation to represent the joint \ac{APP} mass function in the linear domain. 
Fig.~\ref{fig:cc_rank_histogram} shows the histogram of~$r_\text{max}$ for the TTDet-sweep algorithm across different $\ebno$ values for the \ac{BCH}$(63,30)$ code.
We can observe that in all cases, the majority of decoding instances require relatively low ranks (note the logarithmic scale of the relative frequency axis in Fig.~\ref{fig:cc_rank_histogram}).
At ${\ebno=1\,}$dB, however, instances with high ranks~${r_\text{max}>1000}$ do occur, with a mean ${r_\text{max}}$ of $322$ and a median of $144$.
As $\ebno$ increases, the occurence of high ranks decreases significantly.
At ${\ebno=4\,}$dB, the mean (median) $r_\text{max}$ reduces to $17$ ($14$), confirming the low-rank nature of the problem, particularly at high $\ebno$.

\section{Conclusion}\label{sec:conclusion}
We propose a \ac{TT}-based framework for tractable Bayesian inference in discrete-input additive noise models, motivated by the observation that the joint log-\ac{APP} mass function often admits an exact low-rank representation in the \ac{TT} format.
To perform the exponentiation in the \ac{TT} format required for marginalization, we employ a truncated Taylor-series initialization followed by a \ac{TT}-cross approximation, and compare two complementary variants: a sample-based algorithm that is more robust at low \ac{SNR}, and a \ac{DMRG}-based scheme that achieves higher accuracy at high \ac{SNR}.
We demonstrated the proposed framework for \ac{MIMO} detection and for decoding binary linear block codes, deriving explicit low-rank \ac{TT} constructions for both problems.
Numerical results confirm near-optimal performance across a wide \ac{SNR} range while maintaining modest \ac{TT} ranks, demonstrating the efficiency and effectiveness of the proposed framework. Moreover, the framework allows for tuning the balance between accuracy and computational complexity. Opting for smaller maximal ranks results in a controlled reduction of accuracy.

Promising directions for future work include exploring alternative approaches to approximate the exponentiation of the log-\ac{APP}.
Another direction is the extension of the framework to hierarchical \acp{TTN}, which may provide even more compact and efficient representations tailored to the structure of the underlying problem.
We highlight the flexibility of the \ac{TT} framework, which enables the direct adaptation of established signal processing techniques to tensor networks, thereby opening new research opportunities and directions.

\section*{Acknowledgments}
We thank Alexander Fengler and Jonathan Mandelbaum for helpful discussions.

\begin{appendix}
\subsection{Example for Tensor Train Decomposition}\label{appendix:ex}
\begin{figure}[t]
    \centering
    \begin{tikzpicture}[scale=1.5, transform shape]
        \def\xstep{0.55}
        \def\ystep{0.55}
        \def\zdx{-0.245}
        \def\zdy{-0.245}

        \pic at (0,0) (c101) {cube2};
        \pic at (\xstep,0) (c111) {cube2};
        \pic at (0,\ystep) (c001) {cube2};
        \pic at (\xstep,\ystep) (c011) {cube2};

        \pic at (\zdx,\zdy) (c) {cube};
        \pic at (\xstep+\zdx,\zdy) (c111) {cube};
        \pic at (\zdx,\ystep+\zdy) (c001) {cube};
        \pic at (\xstep+\zdx,\ystep+\zdy) (c011) {cube};

        \node[scale=0.6667] at (-0.6,0)(A){$A=$};
        
        \node[scale=0.6667, kit-orange] at (1.7,\zdy/2) (A1){$A(:,2,:)$};
        \node[scale=0.6667, kit-cyan] at (1.7-\zdx,-\zdy/2) (A2){$A(:,1,:)$};

        \draw[kit-orange, thick, dotted] ($(A1.south west)+(0.02,0.15)$) -- ($(A1.south west)+(-0.26,0.15)$);
        \draw[kit-cyan, thick, dotted] ($(A2.south west)+(0.02,0.15)$) -- ($(A2.south west)+(-0.26,0.15)$);
    \end{tikzpicture}
    \caption{Visualization of a tensor~$A\in \R^{2\times 2 \times 2}$ as defined in~\eqref{app:eq:slices}. Each cube represents one tensor entry. The slices $A(:,1,:)$ and $A(:,2,:)$ correspond to $2 \times 2$ matrices, colored in blue and orange, respectively.}
    \label{fig:tensor_illustrate}
\end{figure}
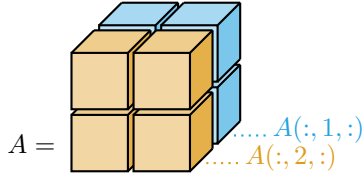

We provide a simple example to build some intuition for the construction of \ac{TT} representations. To illustrate the idea of a low-rank decomposition, consider the tensor $A\in \R^{2\times 2 \times 2}$ with entries
\begin{align}
    A(:,1,:) = \begin{pmatrix}
        1 & 2 \\ 2 & 4
    \end{pmatrix}, \quad A(:,2,:) = \begin{pmatrix}
        2 & 4 \\ 4 & 8
    \end{pmatrix}, \label{app:eq:slices}
\end{align}
as visualized in Fig.~\ref{fig:tensor_illustrate}. 

A trivial \ac{TT} representation of~$A$ is obtained using the core tensors
\begin{align*}
    G_1 &= \begin{pmatrix}
        1 & 0 \\ 0 & 1
    \end{pmatrix} \in \R^{1 \times 2 \times 2},\\ 
    G_2 &= A \in \R^{2 \times 2 \times 2}, \\
    G_3 &= \begin{pmatrix}
        1 & 0 \\ 0 & 1
    \end{pmatrix} \in \R^{2 \times 2 \times 1}.
\end{align*}
As described in Sec.~\ref{sec:TT}, we can reconstruct individual entries of the full tensor by multiplying the corresponding slices of each core tensors. For example, to compute $A(1,2,2)$, we select the slices
\begin{align*}
    G_1(:,1,:) &= \begin{pmatrix}
        1 & 0
    \end{pmatrix}, \quad  G_2(:,2,:) = \begin{pmatrix}
        2 & 4 \\ 4 & 8
    \end{pmatrix}, \\
    G_3(:,2,:) &= \begin{pmatrix}
        0 \\ 1
    \end{pmatrix}.
\end{align*}
and obtain
\begin{align*}
    A(1,2,2) = \begin{pmatrix}
        1 & 0
    \end{pmatrix} \begin{pmatrix}
        2 & 4 \\ 4 & 8
    \end{pmatrix} \begin{pmatrix}
        0 \\ 1
    \end{pmatrix} = 4.
\end{align*}
This representation does not yield any compression, as it essentially reproduces the original tensor. 
However, the tensor~$A$ exhibits significant linear dependencies: both slices $A(:,1,:)$ and $A(:,2,:)$ have linearly dependent columns, and $A(:,2,:)$ is a scaled version of $A(:,1,:)$. 
Hence, the tensor contains redundant information. 

The key idea of a low-rank tensor decomposition is to exploit and remove such redundancies. 
In this example, tensor~$A$ admits an exact rank–$1$ \ac{TT} representation with the core tensors
\begin{align*}
    G_i = \begin{pmatrix}
        1 \\ 2
    \end{pmatrix} \in \R^{1\times 2 \times 1}, \quad i=1,2,3.
\end{align*}
This representation is exact while requiring fewer parameters compared to the full tensor. 

\subsection{Intuition behind the \ac{TT}-Cross Algorithm} \label{app:cross}
We briefly outline the main idea underlying cross approximation algorithms.
First, we explain the concept for matrices and then discuss how it generalizes to tensors and the \ac{TT} format.

\begin{figure}
    \centering
    \tikzset{
    col/.style={
        draw=black,
        fill=kit-cyan!60,
        rounded corners=2pt,
        line width=1pt,
    },
    row/.style={
        draw=black,
        fill=kit-orange,
        fill opacity=0.65,
        rounded corners=2pt,
        line width=1pt,
    },
    core/.style={
        draw=black,
        fill=kit-orange!65!kit-cyan!60,
        rounded corners=1pt,
        line width=1pt,
    },
}
    $\begin{pmatrix}
    \begin{tikzpicture}
        \draw[col] (0.3,-0.1) rectangle (0.5,1.8);
        \draw[col] (0.6,-0.1) rectangle (0.8,1.8);
        \draw[col] (1.2,-0.1) rectangle (1.4,1.8);

        \draw[row] (-0.1,0.3) rectangle (1.8,0.5);
        \draw[row] (-0.1,0.6) rectangle (1.8,0.8);
        \draw[row] (-0.1,1.2) rectangle (1.8,1.4);

    \fill[pattern={Lines[angle=45, distance=1.3pt]}, pattern color=kit-green] (0.315,0.315) rectangle (0.485,0.485);
    \fill[pattern={Lines[angle=45, distance=1.3pt]}, pattern color=kit-green] (0.615,0.315) rectangle (0.785,0.485);
    \fill[pattern={Lines[angle=45, distance=1.3pt]}, pattern color=kit-green] (1.215,0.315) rectangle (1.385,0.485);

    \fill[pattern={Lines[angle=45, distance=1.3pt]}, pattern color=kit-green] (0.315,0.615) rectangle (0.485,0.785);
    \fill[pattern={Lines[angle=45, distance=1.3pt]}, pattern color=kit-green] (0.615,0.615) rectangle (0.785,0.785);
    \fill[pattern={Lines[angle=45, distance=1.3pt]}, pattern color=kit-green] (1.215,0.615) rectangle (1.385,0.785);

    \fill[pattern={Lines[angle=45, distance=1.3pt]}, pattern color=kit-green] (0.315,1.215) rectangle (0.485,1.385);
    \fill[pattern={Lines[angle=45, distance=1.3pt]}, pattern color=kit-green] (0.615,1.215) rectangle (0.785,1.385);
    \fill[pattern={Lines[angle=45, distance=1.3pt]}, pattern color=kit-green] (1.215,1.215) rectangle (1.385,1.385);
    
    \end{tikzpicture}
    \end{pmatrix} =
    \begin{pmatrix}
    \begin{tikzpicture}
        \draw[row, draw=white,fill=white] (-0.05,0) rectangle (0.85,0.2);
    
        \draw[col] (0,-0.1) rectangle (0.2,1.8);
        \draw[col] (0.3,-0.1) rectangle (0.5,1.8);
        \draw[col] (0.6,-0.1) rectangle (0.8,1.8);
    \end{tikzpicture}
    \end{pmatrix}
    \begin{pmatrix}
    \begin{tikzpicture}
        \draw[core] (0,0) rectangle (0.2,0.2);
        \draw[core] (0.3,0) rectangle (0.5,0.2);
        \draw[core] (0.6,0) rectangle (0.8,0.2);

        \draw[core] (0,0.3) rectangle (0.2,0.5);
        \draw[core] (0.3,0.3) rectangle (0.5,0.5);
        \draw[core] (0.6,0.3) rectangle (0.8,0.5);

        \draw[core] (0,0.6) rectangle (0.2,0.8);
        \draw[core] (0.3,0.6) rectangle (0.5,0.8);
        \draw[core] (0.6,0.6) rectangle (0.8,0.8);

        \fill[pattern={Lines[angle=45, distance=1.3pt]}, pattern color=kit-green] (0.015,0.015) rectangle 
        (0.185,0.185);
        \fill[pattern={Lines[angle=45, distance=1.3pt]}, pattern color=kit-green] (0.315,0.015) rectangle 
        (0.485,0.185);
        \fill[pattern={Lines[angle=45, distance=1.3pt]}, pattern color=kit-green] (0.615,0.015) rectangle 
        (0.785,0.185);
        
        \fill[pattern={Lines[angle=45, distance=1.3pt]}, pattern color=kit-green] (0.015,0.315) rectangle 
        (0.185,0.485);
        \fill[pattern={Lines[angle=45, distance=1.3pt]}, pattern color=kit-green] (0.315,0.315) rectangle 
        (0.485,0.485);
        \fill[pattern={Lines[angle=45, distance=1.3pt]}, pattern color=kit-green] (0.615,0.315) rectangle 
        (0.785,0.485);

        \fill[pattern={Lines[angle=45, distance=1.3pt]}, pattern color=kit-green] (0.015,0.615) rectangle 
        (0.185,0.785);
        \fill[pattern={Lines[angle=45, distance=1.3pt]}, pattern color=kit-green] (0.315,0.615) rectangle 
        (0.485,0.785);
        \fill[pattern={Lines[angle=45, distance=1.3pt]}, pattern color=kit-green] (0.615,0.615) rectangle 
        (0.785,0.785);

        \draw[core,fill=none] (0,0) rectangle (0.2,0.2);
        \draw[core,fill=none] (0.3,0) rectangle (0.5,0.2);
        \draw[core,fill=none] (0.6,0) rectangle (0.8,0.2);

        \draw[core,fill=none] (0,0.3) rectangle (0.2,0.5);
        \draw[core,fill=none] (0.3,0.3) rectangle (0.5,0.5);
        \draw[core,fill=none] (0.6,0.3) rectangle (0.8,0.5);

        \draw[core,fill=none] (0,0.6) rectangle (0.2,0.8);
        \draw[core,fill=none] (0.3,0.6) rectangle (0.5,0.8);
        \draw[core,fill=none] (0.6,0.6) rectangle (0.8,0.8);
    \end{tikzpicture}
    \end{pmatrix}^{-1} \hspace{-0.3em}
    \begin{pmatrix}
    \begin{tikzpicture}
        \draw[row] (-0.1,0) rectangle (1.8,0.2);
        \draw[row] (-0.1,0.3) rectangle (1.8,0.5);
        \draw[row] (-0.1,0.6) rectangle (1.8,0.8);
    \end{tikzpicture}
    \end{pmatrix}
    $
    \caption{Graphical illustration of the cross decomposition in~\eqref{eq:Decomp_TT_cross}.}
    \label{fig:cross_explain}
\end{figure}
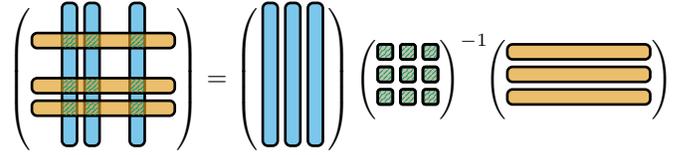

Let ${A\in\R^{n \times m}}$, let $f$ be an element-wise function, and let ${B=f(A)}$. 
Suppose that $B$ is of low rank ${r<n}$. Then there exist $r$ linearly independent columns of $B$, whose indices we collect in the set~${\mathcal{J}=(j_1,\dots,j_r)}$. While $\mathcal{J}$ is assumed to be known here for simplicity, it is determined in practice via an iterative alternating-directions search. 
Since $r$ is small, we can compute the matrix ${C=B(:,\mathcal{J}) \in \R^{n \times r}}$ by evaluating only $nr$ entries of the full matrix ${B=f(A)}$. We then choose $r$ rows of $C$. Specifically, we select the row index set ${\mathcal{I}=(i_1,\dots,i_r)}$ such that $|\det (B(\mathcal{I},\mathcal{J}))|$ is maximized, which promotes linear independence and is accomplished efficiently by the \texttt{maxvol} algorithm~\cite{maxvol_algo}. 
This yields the low-rank decomposition
\begin{align}
    B = CB(\mathcal{I},\mathcal{J})^{-1}R, \label{eq:Decomp_TT_cross}
\end{align} 
where $C=B(:,\mathcal{J})$ and $R=B(\mathcal{I},:)$. Because the index sets $\mathcal{I}$ and $\mathcal{J}$ form a cross-shaped structure in the matrix $B$, as visualized in Fig~\ref{fig:cross_explain}, methods of this type are referred to as cross algorithms.

Tensor-cross algorithms generalize the matrix cross decomposition by successively applying this interpolation idea along each physical dimension to obtain a full \ac{TT} decomposition.
To formalize this, let $\textbf{Mat}_j(B)$ denote the matricization of a tensor ${B \in \R^{n_1 \times \dots \times n_N}}$, in which the $k$th row aligns all entries of $B$ whose $j$th index equals $k$. 
To compute the first \ac{TT} core, we apply the matrix cross-decomposition described above to ${\mathbf{Mat}_1(B) \in \mathbb{R}^{n_1 \times m}}$ with ${m = \prod_{i=2}^{N} n_i}$.
Since $m$ is generally too large to store $R = B(\mathcal{I}, :)$ explicitly, we instead set ${G_1 = C \, B(\mathcal{I}, \mathcal{J})^{-1}}$ as the first core tensor. 
The remaining factor $R$ can be reshaped into a tensor of size $r_1 n_2 \times n_3 \times \cdots \times n_N$ by concatenating the indices in $\mathcal{I}$ with the second physical dimension. This procedure is applied recursively until all cores are obtained. A detailed description is given in~\cite[Sec.~3]{oseledets_TT_cross}.

\end{appendix}


\end{document}

%% file: kit_colors.tex
\definecolor{kit-green}{RGB}{0, 150, 130}
\colorlet{kit-green100}{kit-green}
\colorlet{kit-green90}{kit-green!90!white}
\colorlet{kit-green80}{kit-green!80!white}
\colorlet{kit-green70}{kit-green!70!white}
\colorlet{kit-green60}{kit-green!60!white}
\colorlet{kit-green50}{kit-green!50!white}
\colorlet{kit-green40}{kit-green!40!white}
\colorlet{kit-green30}{kit-green!30!white}
\colorlet{kit-green25}{kit-green!25!white}
\colorlet{kit-green20}{kit-green!20!white}
\colorlet{kit-green15}{kit-green!15!white}
\colorlet{kit-green10}{kit-green!10!white}
\colorlet{kit-green5}{kit-green!5!white}

\definecolor{kit-blue}{RGB}{70, 100, 170}
\colorlet{kit-blue100}{kit-blue}
\colorlet{kit-blue90}{kit-blue!90!white}
\colorlet{kit-blue80}{kit-blue!80!white}
\colorlet{kit-blue70}{kit-blue!70!white}
\colorlet{kit-blue60}{kit-blue!60!white}
\colorlet{kit-blue50}{kit-blue!50!white}
\colorlet{kit-blue40}{kit-blue!40!white}
\colorlet{kit-blue30}{kit-blue!30!white}
\colorlet{kit-blue25}{kit-blue!25!white}
\colorlet{kit-blue20}{kit-blue!20!white}
\colorlet{kit-blue15}{kit-blue!15!white}
\colorlet{kit-blue10}{kit-blue!10!white}
\colorlet{kit-blue5}{kit-blue!5!white}

\definecolor{kit-royalblue}{RGB}{0, 45, 76}
\colorlet{kit-royalblue100}{kit-royalblue}
\colorlet{kit-royalblue90}{kit-royalblue!90!white}
\colorlet{kit-royalblue80}{kit-royalblue!80!white}
\colorlet{kit-royalblue70}{kit-royalblue!70!white}
\colorlet{kit-royalblue60}{kit-royalblue!60!white}
\colorlet{kit-royalblue50}{kit-royalblue!50!white}
\colorlet{kit-royalblue40}{kit-royalblue!40!white}
\colorlet{kit-royalblue30}{kit-royalblue!30!white}
\colorlet{kit-royalblue25}{kit-royalblue!25!white}
\colorlet{kit-royalblue20}{kit-royalblue!20!white}
\colorlet{kit-royalblue15}{kit-royalblue!15!white}
\colorlet{kit-royalblue10}{kit-royalblue!10!white}
\colorlet{kit-royalblue5}{kit-royalblue!5!white}

\definecolor{kit-iceblue100}{RGB}{30, 53, 69}
\definecolor{kit-iceblue70}{RGB}{68, 94, 111}
\definecolor{kit-iceblue50}{RGB}{168, 185, 196}
\definecolor{kit-iceblue30}{RGB}{218, 225, 230}

\definecolor{kit-red}{RGB}{162, 34, 35}
\colorlet{kit-red100}{kit-red}
\colorlet{kit-red90}{kit-red!90!white}
\colorlet{kit-red80}{kit-red!80!white}
\colorlet{kit-red70}{kit-red!70!white}
\colorlet{kit-red60}{kit-red!60!white}
\colorlet{kit-red50}{kit-red!50!white}
\colorlet{kit-red40}{kit-red!40!white}
\colorlet{kit-red30}{kit-red!30!white}
\colorlet{kit-red25}{kit-red!25!white}
\colorlet{kit-red20}{kit-red!20!white}
\colorlet{kit-red15}{kit-red!15!white}
\colorlet{kit-red10}{kit-red!10!white}
\colorlet{kit-red5}{kit-red!5!white}

\definecolor{kit-yellow}{RGB}{252, 229, 0}
\colorlet{kit-yellow100}{kit-yellow}
\colorlet{kit-yellow90}{kit-yellow!90!white}
\colorlet{kit-yellow80}{kit-yellow!80!white}
\colorlet{kit-yellow70}{kit-yellow!70!white}
\colorlet{kit-yellow60}{kit-yellow!60!white}
\colorlet{kit-yellow50}{kit-yellow!50!white}
\colorlet{kit-yellow40}{kit-yellow!40!white}
\colorlet{kit-yellow30}{kit-yellow!30!white}
\colorlet{kit-yellow25}{kit-yellow!25!white}
\colorlet{kit-yellow20}{kit-yellow!20!white}
\colorlet{kit-yellow15}{kit-yellow!15!white}
\colorlet{kit-yellow10}{kit-yellow!10!white}
\colorlet{kit-yellow5}{kit-yellow!5!white}

\definecolor{kit-orange}{RGB}{223, 155, 27}
\colorlet{kit-orange100}{kit-orange}
\colorlet{kit-orange90}{kit-orange!90!white}
\colorlet{kit-orange80}{kit-orange!80!white}
\colorlet{kit-orange70}{kit-orange!70!white}
\colorlet{kit-orange60}{kit-orange!60!white}
\colorlet{kit-orange50}{kit-orange!50!white}
\colorlet{kit-orange40}{kit-orange!40!white}
\colorlet{kit-orange30}{kit-orange!30!white}
\colorlet{kit-orange25}{kit-orange!25!white}
\colorlet{kit-orange20}{kit-orange!20!white}
\colorlet{kit-orange15}{kit-orange!15!white}
\colorlet{kit-orange10}{kit-orange!10!white}
\colorlet{kit-orange5}{kit-orange!5!white}

\definecolor{kit-lightgreen}{RGB}{140, 182, 60}
\colorlet{kit-lightgreen100}{kit-lightgreen}
\colorlet{kit-lightgreen90}{kit-lightgreen!90!white}
\colorlet{kit-lightgreen80}{kit-lightgreen!80!white}
\colorlet{kit-lightgreen70}{kit-lightgreen!70!white}
\colorlet{kit-lightgreen60}{kit-lightgreen!60!white}
\colorlet{kit-lightgreen50}{kit-lightgreen!50!white}
\colorlet{kit-lightgreen40}{kit-lightgreen!40!white}
\colorlet{kit-lightgreen30}{kit-lightgreen!30!white}
\colorlet{kit-lightgreen25}{kit-lightgreen!25!white}
\colorlet{kit-lightgreen20}{kit-lightgreen!20!white}
\colorlet{kit-lightgreen15}{kit-lightgreen!15!white}
\colorlet{kit-lightgreen10}{kit-lightgreen!10!white}
\colorlet{kit-lightgreen5}{kit-lightgreen!5!white}

\definecolor{kit-purple}{RGB}{163, 16, 124}
\colorlet{kit-purple100}{kit-purple}
\colorlet{kit-purple90}{kit-purple!90!white}
\colorlet{kit-purple80}{kit-purple!80!white}
\colorlet{kit-purple70}{kit-purple!70!white}
\colorlet{kit-purple60}{kit-purple!60!white}
\colorlet{kit-purple50}{kit-purple!50!white}
\colorlet{kit-purple40}{kit-purple!40!white}
\colorlet{kit-purple30}{kit-purple!30!white}
\colorlet{kit-purple25}{kit-purple!25!white}
\colorlet{kit-purple20}{kit-purple!20!white}
\colorlet{kit-purple15}{kit-purple!15!white}
\colorlet{kit-purple10}{kit-purple!10!white}
\colorlet{kit-purple5}{kit-purple!5!white}

\definecolor{kit-brown}{RGB}{167, 130, 46}
\colorlet{kit-brown100}{kit-brown}
\colorlet{kit-brown90}{kit-brown!90!white}
\colorlet{kit-brown80}{kit-brown!80!white}
\colorlet{kit-brown70}{kit-brown!70!white}
\colorlet{kit-brown60}{kit-brown!60!white}
\colorlet{kit-brown50}{kit-brown!50!white}
\colorlet{kit-brown40}{kit-brown!40!white}
\colorlet{kit-brown30}{kit-brown!30!white}
\colorlet{kit-brown25}{kit-brown!25!white}
\colorlet{kit-brown20}{kit-brown!20!white}
\colorlet{kit-brown15}{kit-brown!15!white}
\colorlet{kit-brown10}{kit-brown!10!white}
\colorlet{kit-brown5}{kit-brown!5!white}

\definecolor{kit-cyan}{RGB}{35, 161, 224}
\colorlet{kit-cyan100}{kit-cyan}
\colorlet{kit-cyan90}{kit-cyan!90!white}
\colorlet{kit-cyan80}{kit-cyan!80!white}
\colorlet{kit-cyan70}{kit-cyan!70!white}
\colorlet{kit-cyan60}{kit-cyan!60!white}
\colorlet{kit-cyan50}{kit-cyan!50!white}
\colorlet{kit-cyan40}{kit-cyan!40!white}
\colorlet{kit-cyan30}{kit-cyan!30!white}
\colorlet{kit-cyan25}{kit-cyan!25!white}
\colorlet{kit-cyan20}{kit-cyan!20!white}
\colorlet{kit-cyan15}{kit-cyan!15!white}
\colorlet{kit-cyan10}{kit-cyan!10!white}
\colorlet{kit-cyan5}{kit-cyan!5!white}

\definecolor{kit-gray}{RGB}{0, 0, 0}
\colorlet{kit-gray100}{kit-gray}
\colorlet{kit-gray90}{kit-gray!90!white}
\colorlet{kit-gray80}{kit-gray!80!white}
\colorlet{kit-gray70}{kit-gray!70!white}
\colorlet{kit-gray60}{kit-gray!60!white}
\colorlet{kit-gray50}{kit-gray!50!white}
\colorlet{kit-gray40}{kit-gray!40!white}
\colorlet{kit-gray30}{kit-gray!30!white}
\colorlet{kit-gray25}{kit-gray!25!white}
\colorlet{kit-gray20}{kit-gray!20!white}
\colorlet{kit-gray15}{kit-gray!15!white}
\colorlet{kit-gray10}{kit-gray!10!white}
\colorlet{kit-gray5}{kit-gray!5!white}